# Performance Bounds and Associated Design Principles for Multi-Cellular Wireless OFDMA Systems (with Detailed Proofs)

Rohit Aggarwal, C. Emre Koksal, and Philip Schniter

Dept. of ECE, The Ohio State University, Columbus, OH 43210.
Email: {aggarwar,koksal,schniter}@ece.osu.edu

*Abstract*—In this paper, we consider the downlink of large-scale multi-cellular OFDMA-based networks and study performance bounds of the system as a function of the number of users $K$, the number of base-stations $B$, and the number of resource-blocks $N$. Here, a resource block is a collection of subcarriers such that all such collections, that are disjoint have associated independently fading channels. We derive novel upper and lower bounds on the sum-utility for a general spatial geometry of base stations, a truncated path loss model, and a variety of fading models (Rayleigh, Nakagami-$m$, Weibull, and LogNormal). We also establish the associated scaling laws and show that, in the special case of fixed number of resource blocks, a grid-based network of base stations, and Rayleigh-fading channels, the sum information capacity of the system scales as $\Theta(B \log \log K/B)$ for extended networks, and as $O(B \log \log K)$ and $\Omega(\log \log K)$ for dense networks. Interpreting these results, we develop some design principles for the service providers along with some guidelines for the regulators in order to achieve provisioning of various QoS guarantees for the end users and, at the same time, maximize revenue for the service providers.

## I. INTRODUCTION

With the widespread usage of smart phones and the increasing demand for numerous mobile applications, wireless cellular networks have grown significantly in size and complexity. Consequently, the decisions regarding the deployment of base stations, the ratio of the number of subscribers to the number of base stations, the amount to be spent on purchasing more bandwidth, and the revenue model to choose have become much more complicated for service providers. Understanding the performance limits of large multicellular networks and the optimal balance between the number of base stations, the number of subscribers, and the bandwidth to achieve those limits are critical components of the decisions made. Given that the most significant fraction of the performance growth of wireless networks in the last few decades is associated [1] with the cell sizes and the amount of available bandwidth, the aforementioned issues become more important.

To that end, in this paper we analyze the achievable downlink information rate in large multicellular OFDMA systems as a function of the number, $K$, of users, the number, $B$, of base-stations, and the number, $N$, of available resource-blocks. Here, a resource block is a collection of subcarriers such that all such collections, that are disjoint have associated independently fading channels. The contributions of this paper can be summarized as follows.

• For a general spatial geometry of the base-stations and the end users, we develop novel upper and lower bounds on the average achievable rate as a function of $K$, $B$, and $N$.

• Then, we consider two asymptotic scenarios in network size: dense networks and extended networks in which (user) nodes have a uniform spatial distribution. We evaluate our bounds for Rayleigh, Nakagami-$m$, Weibull, and LogNormal fading models along with a truncated path-loss model. To evaluate the bounds, we utilize various results from the *extreme value theory*. We also specify the associated scaling laws in all parameters.

• With the developed bounds we consider four different scenarios. In the first scenario, we consider a *femtocell network* and develop an asymptotic condition for $K$, $B$, and $N$ to guarantee a non-diminishing rate for each user. In the second and third scenarios, we consider extended multicell networks and derive bounds for the choice of $K/B$, i.e., the ratio of the number of users to the number of base stations, in order for the service provider to maximize the revenue per base station and at the same time keep the per-user rate above a certain limit. We analyze two different revenue models: for the overall service the users are charged (1) on the number of bits they are served; (2) a constant amount. Finally, we consider an extended multicell network and develop asymptotic conditions for $K$, $B$, and $N$ to guarantee a minimum return on investment for the service provider.

Calculation of achievable performance of wireless networks has been a challenging, and yet an extremely popular problem in the literature. The performance of large networks have been mainly analyzed in the asymptotic regimes and the results have been in the form of scaling laws [2]–[11] following the seminal work by Gupta and Kumar [2]. Unlike these studies, our main bounds are not asymptotic and we take into account both a distance based power-attenuation law and fading into account in our model. Scaling laws for channel models incorporating, both, distance based power-attenuation and fading have been considered in [12], [13]. We assumed a truncated path-loss model unlike these works, which assume an unbounded path-loss model. Unbounded path loss models affect the asymptotic behavior of the achievable rates significantly. For instance, the capacity scaling law of $\Theta(\log K)$ found in [12] arises by exploiting infinite channel-gain of the users close to the base-station, whereas, without path-loss the scaling law changes to $\Theta(\log\log K)$. Further, our analyses take take into account the bandwidth and number of base-stations in large networks.

## II. SYSTEM MODEL

We consider a multi-cellular time-slotted OFDMA-based downlink network with $B$ base-stations (BS) and $K$ active users as shown in Fig. 1. The base-stations lie in a disc of radius $p - R$ ($p > R > 0$), and their locations are arbitrary and deterministic. The users lie in disc of radius $p$ and their locations are uniformly distributed within this disk. We denote the coordinates of BS $i$ ($1 \leq i \leq B$) with $(a_i, b_i)$, and the coordinates of user $k$ ($1 \leq k \leq K$) with $(x_k, y_k)$.

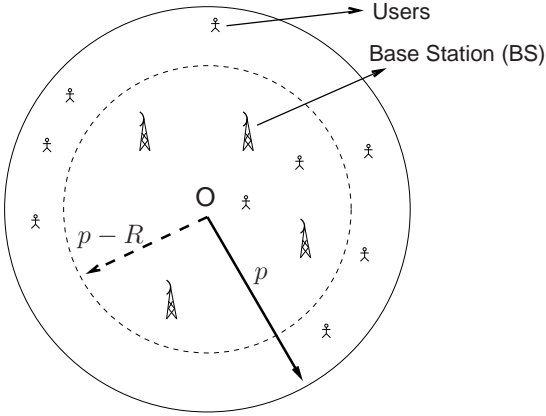

Fig. 1: OFDMA downlink system with $K$ users and $B$ base-stations.

We now describe the channel model. We assume that the OFDMA subchannels are grouped into resource blocks, across which the BSs schedule users for downlink transmission. We denote the complex-valued channel gain over resource-block $n$ ($1 \leq n \leq N$) between user $k$ and BS $i$ by $h_{i,k}^n$. We assume that $h_{i,k}^n$ is composed of the following factors:

$$h_{i,k}^n = \beta R_{i,k}^{-\alpha} \nu_{i,k}^n, \quad (1)$$

where $\beta R_{i,k}^{-\alpha}$ denotes the path-loss attenuation,

$$R_{i,k} = \max\{r_0, \sqrt{(x_k - a_i)^2 + (y_k - b_i)^2}\}, \quad (2)$$

$\beta$, $\alpha$ ($> 1$), $r_0$ ($< R$) are constants, and the fading factor $\nu_{i,k}^n$ is a complex-valued random variable that is i.i.d. across $i, k, n$. Therefore, for any given user-BS combination, the channel-gain over each resource-block is independent[1] of the channel-gains over other resource-blocks [14]. Currently, we keep the distribution of $\nu_{i,k}^n$ general and specific scenarios will be studied in subsequent sections. Assuming unit-variance AWGN, the channel-SNR between user $k$ and BS $i$ across resource-block $n$ can now be defined as

$$\gamma_{i,k}^n \triangleq |h_{i,k}^n|^2 = \beta^2 R_{i,k}^{-2\alpha} |\nu_{i,k}^n|^2. \quad (3)$$

[1] Note that this assumption is for simplicity and our analyses can be easily generalized to the case with dependencies across resource-blocks.

We assume the base-stations have perfect coordination among each other via a backhaul network and they allocate resource-blocks and downlink powers in each time-slot jointly such that the achievable sum-rate of the system is maximized. We denote the corresponding (sum-rate maximizing) user scheduled by BS $i$ across resource-block $n$ by $U_i^n$ and the corresponding (sum-rate maximizing) power allocated by $P_i^n$. We assume that, in each time-slot, every BS allocates powers to scheduled users subject to a sum-power constraint $P_{\text{con}}$. Hence, the set of feasible user allocations and power allocations are:

$$\mathcal{U} \triangleq \{u_i^n \ \forall \ i, n : 1 \leq u_i^n \leq K \ \forall \ i, n\}, \text{ and}$$
$$\mathcal{P} \triangleq \{p_i^n \ \forall \ i, n : \sum_n p_i^n \leq P_{\text{con}} \ \forall \ i\}. \quad (4)$$

Assuming the availability of perfect channel-state information of all users' channel-gains at every BS, the achievable sum-rate of the system can be written as

$$\mathcal{C}_{\boldsymbol{x},\boldsymbol{y},\boldsymbol{\nu}}(\boldsymbol{U},\boldsymbol{P})$$
$$\triangleq \sum_{i=1}^B \sum_{n=1}^N \log\left(1 + \frac{\gamma_{i,U_i^n}^n P_i^n}{1 + \sum_{j \neq i} \gamma_{j,U_i^n}^n P_j^n}\right), \quad (5)$$
$$= \max_{\boldsymbol{u} \in \mathcal{U}, \boldsymbol{p} \in \mathcal{P}} \sum_{i=1}^B \sum_{n=1}^N \log\left(1 + \frac{\gamma_{i,u_i^n}^n p_i^n}{1 + \sum_{j \neq i} \gamma_{j,u_i^n}^n p_j^n}\right), \quad (6)$$

where $\boldsymbol{x} := \{x_k \ \forall \ k\}$, $\boldsymbol{y} := \{y_k \ \forall \ k\}$, $\boldsymbol{\nu} := \{\nu_{i,k}^n \ \forall \ i, k, n\}$, $\boldsymbol{U} := \{U_i^n \ \forall \ i, n\}$, and $\boldsymbol{P} := \{P_i^n \ \forall \ i, n\}$.

Note that $\mathcal{C}_{\boldsymbol{x},\boldsymbol{y},\boldsymbol{\nu}}(\boldsymbol{U},\boldsymbol{P})$ is a random variable, which is a function of the BS-locations $(a_i, b_i)$ for all $i$ (deterministic), and the random variables $(x_k, y_k)$ and $\nu_{i,k}^n$ for all $(i, k, n)$. In the following section, we derive novel upper and lower bounds on the mean of $\mathcal{C}_{\boldsymbol{x},\boldsymbol{y},\boldsymbol{\nu}}(\boldsymbol{U},\boldsymbol{P})$ to determine the scaling laws and develop our network-design guidelines. To state the scaling laws, we use the following notations. For two non-negative functions $f(t)$ and $g(t)$:

1) $f(t) = O(g(t))$ means that there exists a positive constant $c_1$ and an real number $r_1$ such that $f(t) \leq c_1 g(t)$ for all $t \geq r_1$.
2) $f(t) = \Omega(g(t))$ means that there exists a positive constant $c_2$ and an real number $r_2$ such that $f(t) \geq c_2 g(t)$ for all $t \geq r_2$. In other words, $g(t) = O(f(t))$.
3) $f(t) = \Theta(g(t))$ means that $f(t) = O(g(t))$ and $f(t) = \Omega(g(t))$. Note that if $\lim_{t \to \infty} \frac{f(t)}{g(t)} = c_3$ for some constant $c_3$, then $f(t) = \Theta(g(t))$. However, vice versa is not always true.

## III. BOUNDS ON ACHIEVED SUM-RATE

In this section, we derive performance bounds for the system model defined in Section II. The expected achievable sum-rate of the system, using (5), can be written as

$$\mathcal{C}^* = \mathrm{E}\left\{\mathcal{C}_{\boldsymbol{x},\boldsymbol{y},\boldsymbol{\nu}}(\boldsymbol{U},\boldsymbol{P})\right\}, \quad (7)$$

where the expectation is effectively over the SNRs $\{\gamma_{i,k}^n, \forall i, k, n\}$.

The following theorem uses extreme-value theory [15] to derive novel upper and lower bounds on (7).

**Theorem 1.** *The expected achievable sum-rate of the system, $\mathcal{C}^*$, can be bounded as follows:*

$$\sum_{i,n} \mathrm{E}\left\{\frac{\log\left(1 + P_{\mathsf{con}}\max_k \gamma_{i,k}^n\right)}{N + P_{\mathsf{con}}\sum_{j\neq i}\gamma_{j,k}^n}\right\}$$
$$\leq \mathcal{C}^* \leq \min\left\{\sum_{i,n} \mathrm{E}\left\{\log\left(1 + P_{\mathsf{con}}\max_k \gamma_{i,k}^n\right)\right\},\right.$$
$$\left. N\sum_i \mathrm{E}\left\{\log\left(1 + \frac{P_{\mathsf{con}}}{N}\max_{n,k}\gamma_{i,k}^n\right)\right\}\right\}. \quad (8)$$

*Proof:* Proof is given in Appendix A. ∎

## IV. APPLICATION OF BOUNDS TO NETWORKS

The bounds in Theorem 1 are quite general and can be applied to a variety of network and channel settings. In the following subsections, we consider the specific cases of dense and regular extended networks, and apply (8) to obtain the performance bounds and associated scaling laws.

### A. Dense Networks

Dense networks contain a large number of base-stations that are distributed over a fixed area. Typically, such networks occur in dense-urban environments and in dense femtocell deployments. Thus, in our system-model, dense network corresponds to the case in which $p$ is fixed and $K$ and $B$ grow. The following theorem builds on Theorem 1 to find bounds on achievable sum-rate of the system for Rayleigh fading channels.

**Theorem 2.** *For dense networks with Rayleigh fading downlink channels, i.e., $\nu_{i,k}^n \sim \mathcal{CN}(0,1)$,*

$$\left(\log(1 + P_{\mathsf{con}}l_K) + O(1)\right)BN f_{\mathsf{lo}}^{\mathsf{DN}}(r, B, N)$$
$$\leq \mathcal{C}^* \leq \left(\log(1 + P_{\mathsf{con}}l_K) + O(1)\right)BN, \quad (9)$$

*where $r > 0$ is a constant, $l_K = \beta^2 r_0^{-2\alpha}\log\frac{Kr_0^2}{p^2}$, $f_{\mathsf{lo}}^{\mathsf{DN}}(r,B,N) = \frac{r^2}{(1+r^2)(N+P_{\mathsf{con}}\beta^2 r_0^{-2\alpha}(\mu+r\sigma)B)}$, and $\mu, \sigma$ are the mean and standard deviation of $|\nu_{i,k}^n|^2$ ($\mu = \sigma = 1$ for Rayleigh fading channels). The following scaling laws result from (9):*

$$\mathcal{C}^* = O(BN \log\log K), \text{ and}$$
$$\mathcal{C}^* = \Omega(\min\{B, N\}\log\log K). \quad (10)$$

*Proof:* The complete detailed proof is provided in [16]. A detailed sketch with three intermediate lemmas (without proof) are given in Appendix B. To summarize, the first lemma, i.e, Lemma 1, uses Cantelli's inequality and Theorem 1 to show that

$$\mathcal{C}^* \geq f_{\mathsf{lo}}^{\mathsf{DN}}(r,B,N)\sum_{i,n}\mathrm{E}\left\{\log\left(1 + P_{\mathsf{con}}\max_k \gamma_{i,k}^n\right)\right\}.$$

The second lemma, i.e, Lemma 2, finds the distribution of channel-SNR $\gamma_{i,k}^n$ under Rayleigh-distributed $|\nu_{i,k}^n|$ and a truncated path-loss model defined in (1). The third lemma, i.e, Lemma 3, uses Lemma 2 and extreme-value theory to show that $(\max_k \gamma_{i,k}^n - l_K)$ converges in distribution to a limiting random variable with a Gumbel type cdf, that is given by

$$\exp(-e^{-xr_0^{2\alpha}/\beta^2}), \ x \in (-\infty, \infty). \quad (11)$$

Thereafter, we use Lemma 1, Lemma 3, and [9, Theorem A.2] in (38)-(49) to obtain the final result. ∎

**Corollary 1.** *If we have $\frac{\log KN}{N} \gg 1$, then a tighter upper bound for Rayleigh-fading channels can be stated as $\mathcal{C}^* = O(BN\log\frac{\log KN}{N})$.*

The proof sketch is given at the end of Appendix B. For detailed proof, see [16]. Note that the condition $\frac{\log KN}{N} \gg 1$ is easy to satisfy since, typically, the number of users scale much faster than the number of resource blocks (particularly when the number of resource blocks are fixed).

Similar results under different fading models are given by the following theorem.

**Theorem 3.** *If $|\nu_{i,k}^n|$ belongs to either Nakagami-$m$, Weibull, or LogNormal family of distributions, then, for dense networks, the scaling laws for the upper bounds are*

*For Nakagami-$(m,w)$:* $\quad \mathcal{C}^* = O(BN\log\log K)$
*For Weibull$(\lambda, t)$:* $\quad \mathcal{C}^* = O(BN\log\log^{\frac{2}{t}} K)$
*For LogNormal$(a,\omega)$:* $\quad \mathcal{C}^* = O(BN\sqrt{\log K}).$

*and the scaling laws for the lower bounds are*

*For Nakagami-$(m,w)$:* $\quad \mathcal{C}^* = \Omega(BNf_{\mathsf{lo}}^{\mathsf{DN}}(r,B,N)\log\log K)$
*For Weibull$(\lambda,t)$:* $\quad \mathcal{C}^* = \Omega(BNf_{\mathsf{lo}}^{\mathsf{DN}}(r,B,N)\log\log^{\frac{2}{t}} K)$
*For LogNormal$(a,\omega)$:* $\quad \mathcal{C}^* = \Omega(BNf_{\mathsf{lo}}^{\mathsf{DN}}(r,B,N)\sqrt{\log K}).$

*Proof:* Proof sketch given in Appendix C. For detailed proof, see [16]. ∎

**Corollary 2.** *If $\frac{\log KN}{N} \gg 1$, $\frac{\log^{\frac{2}{t}} KN}{N} \gg 1$, and $\frac{e^{\sqrt{\log KN}}}{N} \gg 1$ for Nakagami-$m$, Weibull, and LogNormal distributions, respectively, then tighter upper bounds on the scaling laws can be found as follows:*

*For Nakagami-$(m,w)$:* $\quad \mathcal{C}^* = O\left(BN\log\frac{\log KN}{N}\right)$
*For Weibull$(\lambda, t)$:* $\quad \mathcal{C}^* = O\left(BN\log\frac{\log^{\frac{2}{t}} KN}{N}\right)$
*For LogNormal$(a,\omega)$:* $\quad \mathcal{C}^* = O\left(BN\log\frac{e^{\sqrt{\log KN}}}{N}\right).$

The proof sketch is given at the end of Appendix C. For detailed proof, see [16].

### B. Regular Extended Networks

In extended networks, the area of the network grows with the number of nodes, keeping node density constant. Here we study *regular* extended networks, in which the base-stations lie on a regular grid. We assume the grids are hexagonal as illustrated in Fig. 2 and the distance between two neighboring base-stations is $2R$. Hence, the radius of the network $p = \Theta(R\sqrt{B})$ for large $B$.

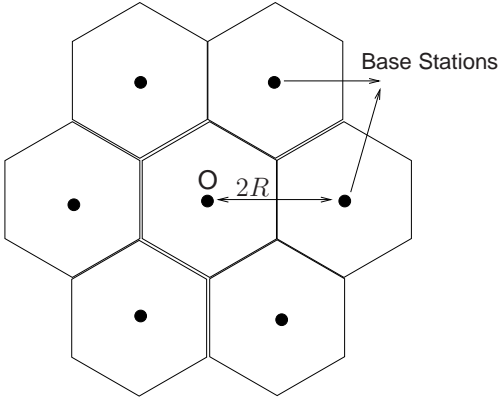

Fig. 2: A regular extended network setup.

The following theorem gives the performance bounds, and the associated scaling laws for regular extended networks and Rayleigh fading channels.

**Theorem 4.** *For regular extended networks with Rayleigh fading downlink channels, i.e., $\nu_{i,k}^n \sim \mathcal{CN}(0,1)$,*

$$\big(\log(1+P_{\mathsf{con}}l_K)+O(1)\big)BNf_{\mathsf{lo}}^{\mathsf{EN}}(r,N)$$
$$\leq \mathcal{C}^* \leq \big(\log(1+P_{\mathsf{con}}l_K)+O(1)\big)BN, \quad (12)$$

*where* $l_K = \beta^2 r_0^{-2\alpha}\log\frac{Kr_0^2}{BR^2}$, $f_{\mathsf{lo}}^{\mathsf{EN}}(r,N) = \frac{(1+r^2)^{-1}r^2}{N+(\mu+r\sigma)c_0}$, $c_0 = \frac{P_{\mathsf{con}}\beta^2 r_0^{2-2\alpha}}{R^2}\Big(4+\frac{\pi}{\sqrt{3}(2\alpha-2)}\Big)$, *and* $\mu,\sigma$ *are the mean and standard deviation of* $|\nu_{i,k}^n|^2$ *($\mu=\sigma=1$ for Rayleigh fading channels), . The associated scaling laws are:*

$$\mathcal{C}^* = O\Big(BN\log\log\frac{K}{B}\Big)$$
$$\mathcal{C}^* = \Omega\Big(B\log\log\frac{K}{B}\Big). \quad (13)$$

*Proof:* Proof sketch is provided in Appendix B. For detailed proof, see [16]. ∎

**Corollary 3.** *If $\frac{\log\frac{KN}{B}}{N} \gg 1$, then a tighter upper bound on the scaling law can be written as $\mathcal{C}^* = O\big(BN\log\frac{\log\frac{KN}{B}}{N}\big)$. This condition is easy to satisfy since, typically, the number of users scale much faster than the number of resource blocks.*

The proof sketch is given at the end of Appendix B. For detailed proof, see [16]. Note that for fixed number of resource blocks $N$, we have from (13),

$$\mathcal{C}^* = \Theta\Big(B\log\log\frac{K}{B}\Big). \quad (14)$$

The following theorem gives similar results for other fading models.

**Theorem 5.** *If $|\nu_{i,k}^n|$ belongs to either Nakagami-$m$, Weibull, or LogNormal family of distributions, then, for regular extended networks, the scaling laws for the upper bounds are:*

*For Nakagami-$(m,w)$:* $\quad \mathcal{C}^* = O\big(BN\log\log\frac{K}{B}\big)$
*For Weibull$(\lambda,t)$:* $\quad \mathcal{C}^* = O\big(BN\log\log^{\frac{2}{t}}\frac{K}{B}\big)$
*For LogNormal$(a,\omega)$:* $\quad \mathcal{C}^* = O\big(BN\sqrt{\log\frac{K}{B}}\big).$

*and the scaling laws for the lower bounds are:*

*For Nakagami-$(m,w)$:* $\quad \mathcal{C}^* = \Omega\big(BNf_{\mathsf{lo}}^{\mathsf{EN}}(r,N)\log\log\frac{K}{B}\big)$
*For Weibull$(\lambda,t)$:* $\quad \mathcal{C}^* = \Omega\big(BNf_{\mathsf{lo}}^{\mathsf{EN}}(r,N)\log\log^{\frac{2}{t}}\frac{K}{B}\big)$
*For LogNormal$(a,\omega)$:* $\quad \mathcal{C}^* = \Omega\big(BNf_{\mathsf{lo}}^{\mathsf{EN}}(r,N)\sqrt{\log\frac{K}{B}}\big).$

*Proof:* Proof sketch is given in Appendix C. For detailed proof, see [16]. ∎

**Corollary 4.** *If $\frac{\log\frac{KN}{B}}{N} \gg 1$, $\frac{\log^{\frac{2}{t}}\frac{KN}{B}}{N} \gg 1$, and $\frac{e^{\sqrt{\log\frac{KN}{B}}}}{N} \gg 1$ for Nakagami, Weibull, and LogNormal distributions, respectively, then tighter upper bounds on the scaling laws can be found as follows:*

*For Nakagami-$(m,w)$:* $\quad \mathcal{C}^* = O\Big(BN\log\frac{\log\frac{KN}{B}}{N}\Big)$
*For Weibull$(\lambda,t)$:* $\quad \mathcal{C}^* = O\Big(BN\log\frac{\log^{\frac{2}{t}}\frac{KN}{B}}{N}\Big)$
*For LogNormal$(a,\omega)$:* $\quad \mathcal{C}^* = O\Big(BN\log\frac{\exp\big\{\sqrt{\log\frac{KN}{B}}\big\}}{N}\Big).$

The proof sketch is given at the end of Appendix C. For detailed proof, see [16].

## V. DESIGN PRINCIPLES

In this section, we interpret the obtained results and outline some fundamental design principles for the service providers and network designers in order to achieve provisioning of various QoS guarantees for the end users and at the same time maximize the revenue for the service providers. For simplicity, we only consider Rayleigh fading channels, though, we note that similar results can be obtained for different fading models. We focus on four scenarios in this paper. In the sequel, we call our system *scalable* under a certain condition, if the condition is not violated as the number of users $K \to \infty$.

**Principle 1.** *In dense femtocell deployments, with the condition that the per-user throughput remains above a certain lower bound, for the system to be scalable, $BN$ must scale as $\Omega\big(\frac{K}{\log\log K}\big)$.*

We use the dense-network abstraction for a dense femtocell deployment [17] where the service operator wants to maintain a minimum throughput per user. In such cases, a necessary condition that the service provider must satisfy is:

$$\frac{BN\Big(\log\Big(1+P_{\mathsf{con}}\beta^2 r_0^{-2\alpha}\log\frac{Kr_0^2}{p^2}\Big)+s\Big)}{K} \geq \bar{s},$$

for some $\bar{s} > 0$, where $s = O(1)$. The above equation implies

$$\frac{BN\log\log K}{K} = \Omega(1). \quad (15)$$

Therefore, the total number of independent resources $BN$, i.e., the product of number of base stations and the number of resource blocks (i.e., the bandwidth), must scale no slower than $\frac{K}{\log\log K}$. Otherwise, then the system is not scalable and a minimum per-user throughput requirement cannot be maintained.

**Principle 2.** *In a large extended multi-cellular network, if the users are charged based on the number of bits they download and there is a unit cost for each base station incurred by the service provider, then there is a finite range of values for the user-density $\frac{K}{B}$ in order to maximize return-on-investment of the service provider while maintaining a minimum per-user throughput.*

Consider a regular extended network with fixed number of resource blocks $N$. In this case, we have $\mathcal{C}^* = \Theta\big(B \log \log \frac{K}{B}\big)$ from (14). We assume a revenue model for the service provider wherein the service provider charges per bit provided to the users. Thus, the overall return on investment of the service provider is proportional to the achievable sum-rate per base-station. In large scale systems (large $K$), the associated optimization problem is:

$$\max_{K,B} c \log\left(1 + P_{\mathsf{con}}\beta^2 r_0^{-2\alpha} \log \frac{Kr_0^2}{BR^2}\right)$$
$$\text{s.t.} \quad \frac{cB \log(1 + P_{\mathsf{con}}\beta^2 r_0^{-2\alpha} \log \frac{Kr_0^2}{BR^2})}{K} \geq \bar{s}, \quad (16)$$

for some $\bar{s} > 0$, where $c$ is a constant bounded as described in (12)-(13). For simplicity of the analysis, let $\beta = r_0 = R = P_{\mathsf{con}} = 1$ and $\frac{\bar{s}}{c} = 0.1$ (in respective SI units). Defining $\rho \triangleq \frac{K}{B}$, the above problem becomes a convex optimization problem in the variable $\rho$. Then, finding the optimal $\rho$ via Lagrange multiplier method, the following Karush-Kuhn-Tucker (KKT) condition must be satisfied:

$$\rho = \frac{(\lambda + 1)10}{(1 + \log \rho)\lambda}, \quad (17)$$

where $\lambda \geq 0$ is the Lagrange multiplier. Note that, the Lagrange multiplier represents the cost associated with violating the per-user throughput constraint. The plots of LHS and RHS of (17) along with the constraint curve are plotted for $\lambda = 0.1, 1, \infty$ in Fig. 3. Here, the constraint curve (see the constraint in (16)) is given by $\frac{c}{\bar{s}} \log(1+\log \rho)$. Note that according to (16), the constraint is satisfied if $\rho \leq \frac{c}{\bar{s}} \log(1 + \log \rho)$, i.e., when the constraint curve lies above the LHS curve. We notice from Fig. 3 that this occurs when $\rho \in [1.1, 12.7]$. Since the optimal solution satisfies the constraint in (16), the optimal $\rho$ lies in the set $[1.1, 12.7]$. Figure 3 also shows that as the cost of violating the constraint, i.e., $\lambda$, increases, the optimal user-density $\rho$ for a given $\lambda$ satisfying (17), i.e., the value of $\rho$ at the intersection point of LHS and RHS curves in Fig. 3, decreases. Since $\rho = 4.1$ corresponds to $\lambda = \infty$, the optimal $\rho$ is greater than or equal to $4.1$. Summarizing the above observations, the optimal user density (number of users per base-station) lies in the closed set $[4.1, 12.7]$, a finite range of values as stated in Principle 2.

To investigate the variation of optimal $\rho$ for a given $\lambda$, denoted by $\rho^*(\lambda)$, we plot $\rho^*(\lambda)$-versus-$\lambda$ in Fig. 4. As shown earlier, the optimal user density lies in a finite range (here, between $4.1$ and $12.7$ users/BS for $\lambda > 0.29$). Furthermore, the optimal user-density $\rho^*(\lambda)$ is a strictly-decreasing convex function of the cost associated with violating the per-user throughput constraint, i.e., $\lambda$.

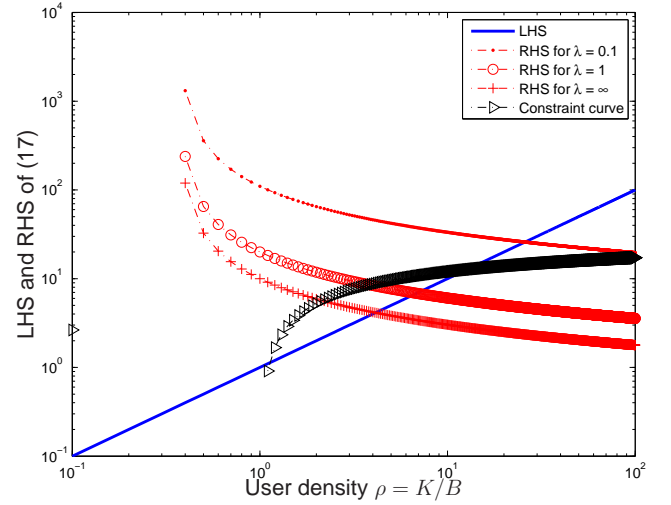

Fig. 3: LHS and RHS of (17) as a function of $\rho$.

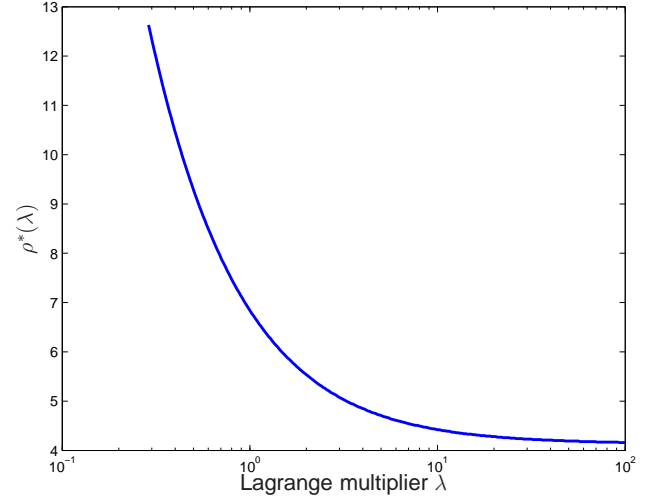

Fig. 4: Optimal user-density, i.e, $\rho^*(\lambda)$, as a function of $\lambda$.

**Principle 3.** *In a large extended multi-cellular network, if the users are charged a fixed amount regardless of the number of bits they download and there is a unit cost for each base station incurred by the service provider, then there is a finite range of values for the user-density $\frac{K}{B}$ in order to maximize return-on-investment of the service provider while maintaining a minimum per-user throughput.*

Consider a regular extended network with fixed number of resource blocks $N$, similar to that assumed in Principle 2. Here, we assume a revenue model for the service provider wherein the service provider charges each user a fixed amount regardless of the number of bits the user downloads. Then, the return on investment of the service provider is proportional to the user-density $\rho = \frac{K}{B}$. In large scale systems (large $K$), the associated optimization problem is:

$$\max_{K,B} s\frac{K}{B}$$
$$\text{s.t.} \quad \frac{cB \log(1 + P_{\text{con}}\beta^2 r_0^{-2\alpha} \log \frac{Kr_0^2}{BR^2})}{K} \geq \bar{s} \quad (18)$$

for some constants $c, s, \bar{s} > 0$. Here, $s$ depends on the amount users are charged by the service provider, and $c$ can be bounded according to (12)-(13). For simplicity of analysis, let $\beta = r_0 = R = P_{\text{con}} = 1$ (in respective SI units). Similar to Principle 2, the above problem becomes a convex optimization problem in the variable $\rho \triangleq \frac{K}{B}$. Therefore, the optimal solution, denoted by $\rho^*$, satisfies the constraint in (18) with equality. In particular, we must satisfy

$$\frac{\bar{s}}{c} \leq \frac{\log(1 + \log \rho)}{\rho}, \quad (19)$$

for all feasible values of $\rho$. The plot of LHS and RHS of (19) as a function of $\rho$ (for $\rho \geq 1$) is plotted in Fig. 5. Examining

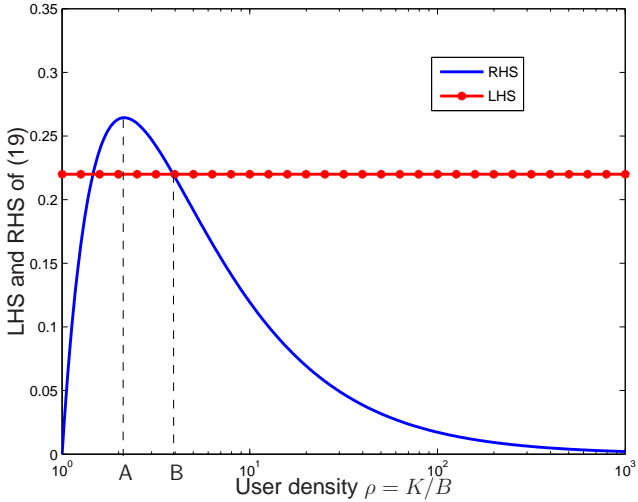

Fig. 5: LHS and RHS of (19) as a function of $\rho$.

(19) and Fig. 5, we note that the per-user throughput constraint is satisfied only if $\frac{\bar{s}}{c} \in [0, 0.26]$. Moreover, for a given value of $\frac{\bar{s}}{c}$, $\rho$ lies in a closed set (for which the RHS curve remains above the LHS curve). The maximum value of $\rho$ in this closed set, i.e., the value of $\rho$ at point $B$ in Fig. 5, is the one that maximizes the objective in (18), i.e., $sK/B$. Hence, it is the optimal $\rho$ for the given value of $\bar{s}/c$. Let us denote it by $\rho^*(\bar{s}/c)$. Note that $\rho^*(\bar{s}/c) \geq 2.14$ (since point $B$ lies to the right of point $A$ in Fig. 5).

If $\bar{s}/c$ is known exactly, then the optimal user-density $\rho^* = \rho^*(\bar{s}/c)$. If not, we can write from (12)-(13) that $c_{\text{lb}} \leq c \leq c_{\text{ub}}$, for some positive constants $c_{\text{lb}}, c_{\text{ub}}$. Then, $\rho^* \in [\rho^*(\bar{s}/c_{\text{lb}}), \rho^*(\bar{s}/c_{\text{ub}})]$. Moreover, since $\rho^*(\bar{s}/c) \geq 2.14$ for all $\bar{s}/c \in [0, 0.26]$, we have $\rho^*(\bar{s}/c_{\text{ub}}) \geq \rho^*(\bar{s}/c_{\text{lb}}) \geq 2.14$.

**Principle 4.** *In regular extended networks, if the users are charged based on the number of bits they download and there is a unit cost for each base station and a cost $c_N$ for unit resource block incurred by the service provider, with the condition that return-on-investment remains above a certain lower bound, then for fixed $B$, the system is scalable only if $N = O(\log K)$, and for fixed $N$, the system is scalable only if $B = O(K)$.*

Consider the case of a regular extended networks with large $K$. Using (8) in conjunction with (40) (using the upper bound obtained via Jensen's inequality in (8)), we have

$$\mathcal{C}^* \leq \left(\log\left(1 + \frac{P_{\text{con}}}{N}l_K + \frac{P_{\text{con}}}{N}\log\log K\right) + O(1)\right)BN$$
$$\approx BN \log\left(\frac{P_{\text{con}}}{N}l_K\right), \text{ for large } \frac{P_{\text{con}}l_K}{N}, \quad (20)$$

where $l_K = \beta^2 r_0^{-2\alpha} \log \frac{KNr_0^2}{BR^2}$. For simplicity of analysis, let $P_{\text{con}} = \beta = r_0 = R = 1$ (in their respective SI units). If the service provider wants to maintain a minimum level of return-on-investment, then

$$\frac{BN}{B + c_N N} \log\left(\frac{1}{N} \log \frac{KN}{B}\right) > \bar{s}, \quad (21)$$

for some $\bar{s} > 0$. The above equation implies

$$N = O(\log K), \text{ for fixed } B, \text{ and}$$
$$B = O(K), \text{ for fixed } N. \quad (22)$$

## VI. CONCLUSION

In this paper, we developed bounds for the achievable downlink rate in multi-cellular OFDMA based networks and specified the associated scaling laws with respect to number of users $K$, number of base-stations $B$, and number of resource-blocks $N$ (bandwidth). Our general bounds hold for a general spatial distribution of base-stations, a truncated path-loss model, and a general channel-fading model. We evaluated the bounds for *dense* and *extended* networks in which nodes were distributed uniformly for Rayleigh, Nakagami-$m$, Weibull, and LogNormal fading models. We showed that for dense networks, under Rayleigh fading channels, the achievable rate is lower bounded by $\Omega(\min(B, N) \log \log K)$, and upper bounded by $O(BN \log \log K)$. The corresponding result for regular extended networks showed that the capacity is lower bounded by $\Omega(B \log \log \frac{K}{B})$ and upper bounded by $O(BN \log \log \frac{K}{B})$. We derived similar results for Nakagami-$m$, Weibull, and LogNormal family of fading models.

We then applied the obtained results to develop four design principles for the service providers and regulators to achieve QoS provisioning along with system scalability. According to the first principle, in dense-femtocell deployments, if a minimum per-user throughput requirement must be maintained, then the system is scalable only if $BN$ scales as $\Omega(\frac{K}{\log \log K})$. In the second and the third principles, we considered two user-charging methods: per-bit and fixed, and showed that the user density must be kept within a finite range of values in order to maximize the return on investment, while keeping the per-user rate above a certain value. Finally, in the fourth principle, we also considered the cost of the bandwidth to the service provider along with the cost of the base stations and showed that for fixed $B$, the system is scalable only if $N = O(\log K)$, and for fixed $N$, the system is scalable only if $B = O(K)$.

## APPENDIX A
## PROOF OF THEOREM 1

By ignoring the interference, we have

$$\mathcal{C}_{\boldsymbol{x},\boldsymbol{y},\boldsymbol{\nu}}(\boldsymbol{U},\boldsymbol{P}) \leq \sum_{i=1}^{B} \sum_{n=1}^{N} \log\left(1 + P_i^n \gamma_{i,U_i(n)}^n\right) \quad (23)$$

This implies

$$\mathcal{C}^* \leq \sum_{i=1}^{B} \sum_{n=1}^{N} \max_k \mathrm{E}\left\{\log\left(1 + P_{\text{con}} \gamma_{i,k}^n\right)\right\}$$

$$\leq \sum_{i=1}^{B} \sum_{n=1}^{N} \mathrm{E}\left\{\max_k \log\left(1 + P_{\text{con}} \gamma_{i,k}^n\right)\right\} \quad (24)$$

$$\leq \sum_{i=1}^{B} \sum_{n=1}^{N} \mathrm{E}\left\{\log\left(1 + P_{\text{con}} \max_k \gamma_{i,k}^n\right)\right\}, \quad (25)$$

where (24) follows because, for any function $f(\cdot,\cdot)$, $\max_k \mathrm{E}\{f(k,\cdot)\} \leq \mathrm{E}\{\max_k f(k,\cdot)\}$, and (25) follows because $\log(\cdot)$ is a non-decreasing function. One can also construct an alternate upper bound by applying Jensen's inequality to the RHS of (23) as follows:

$$\mathcal{C}_{\boldsymbol{x},\boldsymbol{y},\boldsymbol{\nu}}(\boldsymbol{U},\boldsymbol{P}) \leq N \sum_{i=1}^{B} \log\left(1 + \frac{1}{N}\sum_n P_i^n \gamma_{i,U_i(n)}^n\right) \quad (26)$$

$$\leq N \sum_{i=1}^{B} \log\left(1 + \frac{P_{\text{con}}}{N} \max_{n,k} \gamma_{i,k}^n\right), \quad (27)$$

since $\sum_n P_i^n \leq P_{\text{con}}$. Therefore,

$$\mathcal{C}^* \leq N \sum_{i=1}^{B} \mathrm{E}\left\{\log\left(1 + \frac{P_{\text{con}}}{N} \max_{n,k} \gamma_{i,k}^n\right)\right\}. \quad (28)$$

Combining (25) and (28), we obtain the upper bound in Theorem 1.

For lower bound, let $P_{\text{con}}/N$ power be allocated to each resource-block by every BS. Then,

$$\mathcal{C}_{\boldsymbol{x},\boldsymbol{y},\boldsymbol{\nu}}(\boldsymbol{U},\boldsymbol{P})$$
$$\geq \sum_{i=1}^{B} \sum_{n=1}^{N} \log\left(1 + \frac{P_{\text{con}} \gamma_{i,k_i(n)}^n}{N + P_{\text{con}} \sum_{j \neq i} \gamma_{j,k_i(n)}^n}\right), \quad (29)$$

where $k_i(n)$ is any other user allocated on subchannel $n$ by BS $i$. Note that, due to sub-optimal power allocation, all user-allocation strategies $\{k_i(n), \forall i, n\}$ achieve a utility that is lower that $\mathcal{C}_{\boldsymbol{x},\boldsymbol{y},\boldsymbol{\nu}}(\boldsymbol{U},\boldsymbol{P})$. To handle (29) easily, we introduce an indicator variable $I_{i,k}^n(\boldsymbol{x},\boldsymbol{y},\boldsymbol{\nu})$ which equals 1 if $k = k_i(n)$, otherwise takes the value 0. Since, each BS $i$ can schedule at-most one user on any resource block $n$ in a given time-slot, we have $\sum_k I_{i,k}^n(\boldsymbol{x},\boldsymbol{y},\boldsymbol{\nu}) = 1 \; \forall \; i, n$. Now, (29) can be re-written as:

$$\mathcal{C}_{\boldsymbol{x},\boldsymbol{y},\boldsymbol{\nu}}(\boldsymbol{U},\boldsymbol{P}) \quad (30)$$
$$\geq \sum_{i=1}^{B} \sum_{n=1}^{N} \sum_{k=1}^{K} I_{i,k}^n(\boldsymbol{x},\boldsymbol{y},\boldsymbol{\nu}) \log\left(1 + \frac{P_{\text{con}} \gamma_{i,k}^n}{N + P_{\text{con}} \sum_{j \neq i} \gamma_{j,k}^n}\right).$$

Taking expectation w.r.t. $(\boldsymbol{x},\boldsymbol{y},\boldsymbol{\nu})$, we get

$$\mathcal{C}^* \geq \sum_{i,n,k} \mathrm{E}\left\{I_{i,k}^n(\boldsymbol{x},\boldsymbol{y},\boldsymbol{z}) \log\left(1 + \frac{P_{\text{con}} \gamma_{i,k}^n}{N + P_{\text{con}} \sum_{j \neq i} \gamma_{j,k}^n}\right)\right\}$$

$$\geq \sum_{i,n,k} \mathrm{E}\left\{I_{i,k}^n(\boldsymbol{x},\boldsymbol{y},\boldsymbol{z}) \frac{\log\left(1 + P_{\text{con}} \gamma_{i,k}^n\right)}{N + P_{\text{con}} \sum_{j \neq i} \gamma_{j,k}^n}\right\}. \quad (31)$$

Here, the last equation holds because for any non-decreasing concave function $V(\cdot)$ (for example, $V(x) = \log(1+x)$) and for all $d_1, d_2 > 0$, we have

$$V(d_1) - V(0) \leq \left[V\left(\frac{d_1}{d_2}\right) - V(0)\right] d_2$$
$$\implies V\left(\frac{d_1}{d_2}\right) \geq \frac{V(d_1) - V(0)}{d_2} + V(0). \quad (32)$$

Now, $\frac{1}{N + P_{\text{con}} \sum_{j \neq i} \gamma_{j,k}^n(y_{j,k}^n)} \leq 1$. Therefore,

$$\mathcal{C}^* \geq \sum_{i,n,k} \mathrm{E}\left\{\frac{I_{i,k}^n(\boldsymbol{x},\boldsymbol{y},\boldsymbol{\nu}) \log\left(1 + P_{\text{con}} \gamma_{i,k}^n\right)}{N + P_{\text{con}} \sum_{j \neq i} \gamma_{j,k}^n}\right\}. \quad (33)$$

To obtain the best lower bound, we now select the user $k_i(n)$ to be the one for which $\gamma_{i,k}^n$ attains the highest value for every combination $(i,n)$, i.e.,

$$I_{i,k}^n(\boldsymbol{x},\boldsymbol{y},\boldsymbol{\nu}) = \begin{cases} 1 & \text{if } k = \arg\max_{k'} \gamma_{i,k'}^n \\ 0 & \text{otherwise.} \end{cases} \quad (34)$$

Using (34) in (33), we get the lower bound in Theorem 1.

## APPENDIX B
## PROOF OF THEOREM 2 AND THEOREM 4

In this proof, Lemma 1, Lemma 2, Lemma 3, and Lemma 4 have been stated without proofs. For detailed proofs, see the technical report [16].

To prove the final result, we will use three lemmas. Lemma 1 gives a new lower bound from (8) using an application of one-sided variant of Chebyshev's inequality (also called Cantelli's inequality).

**Lemma 1.** *In a dense-network and Rayleigh fading channel, i.e., $\nu_{i,k}^n \sim \mathcal{CN}(0,1)$, the lower bound on the achievable sum-rate of the system is*

$$\mathcal{C}^* \quad (35)$$
$$\geq f_{\text{lo}}^{\text{DN}}(r, B, N) \sum_{i,n} \mathrm{E}_{\boldsymbol{x},\boldsymbol{y},\boldsymbol{\nu}}\left\{\log\left(1 + P_{\text{con}} \max_k \gamma_{i,k}^n\right)\right\},$$

*where $r > 0$ is a fixed number, and $f_{\text{lo}}^{\text{DN}}(r, B, N) = \frac{r^2}{(1+r^2)(N + P_{\text{con}} \beta^2 r_0^{-2\alpha}(\mu + r\sigma)B)}$, and $\mu = \sigma = 1$ for Rayleigh fading channels.*

Lemma 1 in conjunction with Theorem 1 shows that both, the upper and lower bounds, on the achievable sum-rate, i.e., $\mathcal{C}^*$, are functions of SNR scaling. To find the SNR-scaling law, we state the following lemmas.

**Lemma 2.** *Under Rayleigh fading, i.e., $\nu_{i,k}^n \sim \mathcal{CN}(0,1)$, the CDF of $\gamma_{i,k}^n$ is*

$$F_{\gamma_{i,k}^n}(\gamma) = 1 - \frac{r_0^2}{p^2} e^{-\frac{\gamma}{\beta^2 r_0^{-2\alpha}}}$$
$$- \frac{1}{\alpha \beta^2 p^2} \int_{\frac{\beta^2}{(p-d)^{2\alpha}}}^{\beta^2 r_0^{-2\alpha}} e^{-\frac{\gamma}{g}} \left(\frac{g}{\beta^2}\right)^{-1-\frac{1}{\alpha}} dg$$
$$+ \int_{\frac{\beta^2}{(p+d)^{2\alpha}}}^{\frac{\beta^2}{(p-d)^{2\alpha}}} \exp(-\gamma/g) ds(g), \quad (36)$$

*where $d = \sqrt{a_i^2 + b_i^2}$, and $s(g)$ equals*

$$\frac{1}{\pi p^2} \left[ \left(\frac{g}{\beta^2}\right)^{-1/\alpha} \cos^{-1}\left(\frac{d^2 + \left(\frac{g}{\beta^2}\right)^{-1/\alpha} - p^2}{2d\left(\frac{g}{\beta^2}\right)^{-1/2\alpha}}\right) \right.$$
$$+ p^2 \cos^{-1}\left(\frac{d^2 + p^2 - \left(\frac{g}{\beta^2}\right)^{-1/\alpha}}{2dp}\right)$$
$$- \frac{1}{2} \sqrt{\left(p+d-\left(\frac{g}{\beta^2}\right)^{-1/2\alpha}\right)\left(p+\left(\frac{g}{\beta^2}\right)^{-1/2\alpha}-d\right)}$$
$$\left. \times \sqrt{\left(d+\left(\frac{g}{\beta^2}\right)^{-1/2\alpha}-p\right)\left(d+p+\left(\frac{g}{\beta^2}\right)^{-1/2\alpha}\right)} \right].$$

Using Lemma 2, we can now determine the scaling of $\max_k \gamma_{i,k}^n$ for a given $(i,n)$ under Rayleigh fading.

**Lemma 3.** *Let $\gamma_{i,k}^n$ be a random variable with a cdf defined in (36). Then, the growth function $\frac{1-F_{\gamma_{i,k}^n}(\gamma)}{f_{\gamma_{i,k}^n}(\gamma)}$ converges to a constant $\beta^2 r_0^{-2\alpha}$ as $\gamma \to \infty$, and $\gamma_{i,k}^n$ belongs to a domain of attraction [15]. Furthermore, the cdf of $(\max_k \gamma_{i,k}^n - l_K)$ converges in distribution to a limiting random variable with a Gumbel type cdf, that is given by*

$$\exp(-e^{-xr_0^{2\alpha}/\beta^2}), \ x \in (-\infty, \infty), \quad (37)$$

*where $l_K$ is such that $F_{\gamma_{i,k}^n}(l_K) = 1 - 1/K$. In particular, $l_K = \beta^2 r_0^{-2\alpha} \log \frac{K r_0^2}{p^2}$.*

Note that the SNR scaling is independent of $(a_i, b_i)$ and depends on $r_0, p$. Now, we use the above three lemmas to prove the final result in Theorem 2 (and Theorem 4). Since the growth function converges to a constant (see Lemma 3), we apply [9, Theorem A.2] giving us:

$$Pr\{l_K - \log \log K \leq \max_k \gamma_{i,k}^n \leq l_K + \log \log K\}$$
$$\geq 1 - O\left(\frac{1}{\log K}\right), \quad (38)$$

where $l_K = \beta^2 r_0^{-2\alpha} \log \frac{K r_0^2}{p^2}$. Now,

$$E\left\{\log\left(1 + P_{\mathsf{con}} \max_k \gamma_{i,k}^n\right)\right\}$$
$$\leq Pr\left\{\max_k \gamma_{i,k}^n \leq l_K + \log \log K\right\}$$
$$\times \log(1 + P_{\mathsf{con}} l_K + P_{\mathsf{con}} \log \log K)$$
$$+ Pr\left\{\max_k \gamma_{i,k}^n > l_K + \log \log K\right\}$$
$$\times \log(1 + P_{\mathsf{con}} \beta^2 r_0^{-2\alpha} K) \quad (39)$$
$$\leq \log(1 + P_{\mathsf{con}} l_K + P_{\mathsf{con}} \log \log K)$$
$$+ O\left(\frac{1}{\log K}\right) \log(1 + P_{\mathsf{con}} \beta^2 r_0^{-2\alpha} K)$$
$$\leq \log(1 + P_{\mathsf{con}} l_K + P_{\mathsf{con}} \log \log K) + O(1), \quad (40)$$

where, in (39), we have used the fact that the sum-rate is bounded above by $\log(1 + P_{\mathsf{con}} \beta^2 r_0^{-2\alpha} K)$. This is because

$$E\left\{\log\left(1 + P_{\mathsf{con}} \max_k \gamma_{i,k}^n\right)\right\}$$
$$\leq E\left\{\log\left(1 + P_{\mathsf{con}} \sum_k \gamma_{i,k}^n\right)\right\}$$
$$\leq \log\left(1 + P_{\mathsf{con}} \sum_k E\{\gamma_{i,k}^n\}\right)$$
$$\leq \log\left(1 + P_{\mathsf{con}} \beta^2 r_0^{-2\alpha} \sum_k E\{|\nu_{i,k}^n|^2\}\right)$$
$$\leq \log\left(1 + P_{\mathsf{con}} \beta^2 r_0^{-2\alpha} K\right). \quad (41)$$

Further, from (38), we have

$$E\left\{\log\left(1 + P_{\mathsf{con}} \max_k \gamma_{i,k}^n\right)\right\} \quad (42)$$
$$\geq \log(1 + P_{\mathsf{con}} l_K - P_{\mathsf{con}} \log \log K)\left(1 - O\left(\frac{1}{\log K}\right)\right).$$

Combining (40) and (42), we have for large $K$,

$$\log(1 + P_{\mathsf{con}} l_K) + O(1) \leq E\left\{\log\left(1 + P_{\mathsf{con}} \max_k \gamma_{i,k}^n\right)\right\}$$
$$\leq \log(1 + P_{\mathsf{con}} l_K) + O(1), \quad (43)$$

Therefore,

$$\lim_{K \to \infty} \frac{E\left\{\log\left(1 + P_{\mathsf{con}} \max_k \gamma_{i,k}^n\right)\right\}}{\log \log K} = 1. \quad (44)$$

By definition of $\Theta(\cdot)$ in Section I, we have,

$$E\left\{\log\left(1 + P_{\mathsf{con}} \max_k \gamma_{i,k}^n\right)\right\} = \Theta(\log \log K). \quad (45)$$

Using the above equation in conjunction with (8) and (35), we have

$$\mathcal{C}^* = O(BN \log \log K), \text{ and}$$
$$\mathcal{C}^* = \Omega(BN f_{\mathsf{lo}}^{\mathsf{DN}}(r, B, N) \log \log K), \quad (46)$$

where $f_{\mathsf{lo}}^{\mathsf{DN}}(r, B, N)$ is defined in Lemma 1.

Now, to prove Corollary 1 and Corollary 3, we use the upper bound in (8) obtained via Jensen's inequality. In particular, we have

$$\mathcal{C}^* \leq N \sum_i E\left\{\log\left(1 + \frac{P_{\mathsf{con}}}{N} \max_{n,k} \gamma_{i,k}^n\right)\right\} \quad (47)$$

$$\lim_{KN \to \infty} \frac{\mathrm{E}\left\{\log\left(1 + \frac{P_{\mathsf{con}}}{N}\max_{n,k}\gamma_{i,k}^n\right)\right\}}{\log\frac{\log KN}{N}} = 1. \quad (48)$$

This implies

$$\mathcal{C}^* = O\Big(BN \log \frac{\log KN}{N}\Big). \quad (49)$$

Note that the above result is only true if $P_{\mathsf{con}} \log \frac{KN}{N} \gg 1$ (or if, $\log \frac{KN}{N} \gg 1$ since $P_{\mathsf{con}}$ is fixed).

To prove Theorem 4 for regular extended networks, we use the following lemma instead of Lemma 1.

**Lemma 4.** *In a regular extended-network and Rayleigh fading channel, i.e., $\nu_{i,k}^n \sim \mathcal{CN}(0,1)$, the lower bound on achievable sum-rate of the system is*

$$\mathcal{C}^* \geq f_{\mathsf{lo}}^{\mathsf{EN}}(r, N) \sum_{i,n} \mathrm{E}\left\{\log\left(1 + P_{\mathsf{con}}\max_k \gamma_{i,k}^n\right)\right\}, \quad (50)$$

*where $r > 0$ is a fixed number, $f_{\mathsf{lo}}^{\mathsf{EN}}(r,N) = \frac{(1+r^2)^{-1}r^2}{N+(\mu+r\sigma)c_0}$, and $c_0 = \frac{P_{\mathsf{con}}\beta^2 r_0^{2-2\alpha}}{R^2}\left(4 + \frac{\pi}{\sqrt{3}(2\alpha-2)}\right)$.*

The rest of the steps in the proof remain same, with the only change of substituting $p^2$ by $R^2 B$ in Lemma 3 and in the subsequent steps (38)-(49).

## Appendix C
### Proof Sketch of Theorem 3 and Theorem 5

For complete proof, see the technical report [16]. The proof comprises of three parts. First, we take the limit of the growth function $h(\gamma) = \frac{1-F_{\gamma_{i,k}^n}(\gamma)}{f_{\gamma_{i,k}^n}(\gamma)}$ as $\gamma \to \infty$ and show that the distribution of $\gamma_{i,k}^n$ lies in the domain of maximal attraction [15]. In particular, we determine if the limiting distribution of $\gamma_{i,k}^n$ is a Fréchet, Weibull, or Gumbel type, and find $l_K$ in each case such that $1 - F_{\gamma_{i,k}^n}(l_K) = \frac{1}{K}$. Second, using the limiting pdf (Fréchet, Weibull, or Gumbel), we show that $\max_k \gamma_{i,k}^n = \Theta(l_K)$ with high probability. This can be proved by following the steps of [9, Theorem 1] for Gumbel type distributions. Finally, we use the same steps as used (39)-(46) to find the scaling laws of $\mathcal{C}^*$. We briefly give the main calculations, for Nakagami-m, Weibull, and LogNormal distributions here.

1) For Nakagami-$(m,w)$ fading, $h(\gamma) \to \frac{w}{m}\beta^2 r_0^{-2\alpha}$ as $\gamma \to \infty$. Hence, $h'(\gamma) \to 0$ and the limiting distribution of $\max_k \gamma_{i,k}^n$ is of Gumbel type. Solving for $l_K$, it can be shows that for large $K$, $l_K = \frac{w\beta^2 r_0^{-2\alpha}}{m}\log \frac{Kr_0^2 m^{m-1}}{p^2\Gamma(m)(w\beta^2 r_0^{-2\alpha})^{m-1}}$.

2) For Weibull-$(\lambda,t)$ fading with scale $\lambda > 0$ and shape $t > 0$, $h(\gamma) = \frac{2(\lambda^2\beta^2 r_0^{-2\alpha})^{t/2}}{t\gamma^{t/2-1}}$. Therefore, $\lim_{\gamma \to \infty} h'(\gamma) = 0$. Consequently, the limiting distribution of $\max_k \gamma_{i,k}^n$ is of Gumbel type. In this case,

$$l_K = \lambda^2 \beta^2 r_0^{-2\alpha} \log^{\frac{2}{t}} \frac{Kr_0^2}{p^2}.$$

3) For LogNormal fading, i.e, $\log \mathcal{N}(a,\omega)$, we have $h(\gamma) \approx \frac{4\omega\gamma}{\log\gamma}$ for large $\gamma$. Therefore, $\lim_{\gamma \to \infty} h'(\gamma) = 0$, and the limiting distribution of $\gamma_{i,k}^n$ is of Gumbel type. In this case, we finally get

$$l_K = \beta^2 r_0^{-2\alpha} e^{\sqrt{8\omega \log \frac{Kr_0^2}{p^2} + \Theta(\log \log K)}}.$$

Note that $p$ is fixed for dense networks and $p \approx r\sqrt{B}$ for regular extended networks. The proofs of Corollary 2 and Corollary 4 follow from the proofs of Corollary 1 and Corollary 3 in (47) - (49) in Appendix B (by using the upper bound on the achievable sum-rate obtained via Jensen's inequality in Theorem 1).

# Detailed proofs

## I. Proof of Theorem 2 and Corollary 1 for Dense Networks

The proof outline is as follows. We first prove three lemmas. The first lemma, i.e, Lemma I.1, uses one-sided variant of Chebyshev's inequality (also called Cantelli's inequality) and Theorem 1 to show that

$$\mathcal{C}^* \geq f_{\text{lo}}^{\text{DN}}(r, B, N) \sum_{i,n} \text{E}\left\{\log\left(1 + P_{\text{con}} \max_k \gamma_{i,k}^n\right)\right\},$$

where $\mathcal{C}^*$ is expected achievable sum-rate of the system. The second lemma, i.e, Lemma I.2, finds the cumulative distribution function (CDF) of channel-SNR, denoted by $F_{\gamma_{i,k}^n}(\cdot)$, under Rayleigh-distributed $|\nu_{i,k}^n|$ and a truncated path-loss model. The third lemma, i.e, Lemma I.3, uses Lemma I.2 and extreme-value theory to show that $(\max_k \gamma_{i,k}^n - l_K)$ converges in distribution to a limiting random variable with a Gumbel type cdf, that is given by

$$\exp(-e^{-xr_0^{2\alpha}/\beta^2}), \ x \in (-\infty, \infty), \tag{I.1}$$

where $F_{\gamma_{i,k}^n}(l_K) = 1 - \frac{1}{K}$. Thereafter, we use Theorem 1, Lemma I.1, Lemma I.3, and [SH05, Theorem A.2] to obtain the final result.

Now, we give details of the full proof.

**Lemma I.1.** *In a dense-network, the expected achievable sum-rate is lower bounded as:*

$$\mathcal{C}^* \geq f_{\text{lo}}^{\text{DN}}(r, B, N) \sum_{i,n} \text{E}\left\{\log\left(1 + P_{\text{con}} \max_k \gamma_{i,k}^n\right)\right\}, \tag{I.2}$$

*where $r > 0$ is a fixed number, $f_{\text{lo}}^{\text{DN}}(r, B, N) = \frac{r^2}{(1+r^2)(N+P_{\text{con}}\beta^2 r_0^{-2\alpha}(\mu+r\sigma)B)}$, $\mu$ and $\sigma$ are the mean and standard-deviation of $|\nu_{i,k}^n|^2$.*

*Proof:* We know that

$$\sum_{j \neq i} \gamma_{j,k}^n = \beta^2 \sum_{j \neq i} R_{j,k}^{-2\alpha} |\nu_{j,k}^n|^2 \leq \beta^2 r_0^{-2\alpha} \sum_{j \neq i} |\nu_{j,k}^n|^2. \tag{I.3}$$

Therefore, the lower bound in Theorem 1 reduces to the following equation.

$$\mathcal{C}^* \geq \sum_{i,n,k} \text{E}\left\{\frac{\max_k \log\left(1 + P_{\text{con}} \gamma_{i,k}^n\right)}{N + P_{\text{con}} \beta^2 r_0^{-2\alpha} \sum_{j \neq i} |\nu_{j,k}^n|^2}\right\}. \tag{I.4}$$

Now, we apply one-sided variant of Chebyshev's inequality (also called Cantelli's inequality) to the term $\sum_{j \neq i} |\nu_{j,k}^n|^2$ in the denominator. By assumption, $|\nu_{i,k}^n|^2$ are i.i.d. across $i, k, n$ with mean $\mu$ and variance $\sigma$. Hence, applying

Cantelli's inequality, We have

$$Pr\Big(\sum_{j\neq i} |\nu_{j,k}^n|^2 > (B-1)(\mu+r\sigma)\Big) \leq \frac{1}{1+r^2}$$

$$\implies Pr\Big(\sum_{j\neq i} |\nu_{j,k}^n|^2 > (\mu+r\sigma)B\Big) \leq \frac{1}{1+r^2} \quad \text{(I.5)}$$

$$\implies Pr\Big(\sum_{j\neq i} |\nu_{j,k}^n|^2 \leq (\mu+r\sigma)B\Big) \geq \frac{r^2}{1+r^2} \quad \text{(I.6)}$$

where $r > 0$ is a fixed number.

Now, we break the expectation in (I.4) into two parts — one with $\sum_{j\neq i} |\nu_{j,k}^n|^2 > (\mu+r\sigma)B$ and other with $\sum_{j\neq i} |\nu_{j,k}^n|^2 \leq (\mu+r\sigma)B$. We then ignore the first part to obtain another lower bound. Therefore, we now have

$$\mathcal{C}^* \geq \sum_{i=1}^{B}\sum_{n=1}^{N} \mathrm{E}\left\{ \frac{\max_k \log\left(1+P_{\mathsf{con}}\gamma_{i,k}^n\right)}{N+(\mu+r\sigma)BP_{\mathsf{con}}\beta^2 r_0^{-2\alpha}} \bigg|_{\sum_{j\neq i}|\nu_{j,k}^n|^2 \leq (\mu+r\sigma)B}\right\} \times Pr\Big(\sum_{j\neq i} |\nu_{j,k}^n|^2 \leq (\mu+r\sigma)B\Big)$$

$$\geq \frac{\frac{r^2}{1+r^2}}{N+(\mu+r\sigma)BP_{\mathsf{con}}\beta^2 r_0^{-2\alpha}} \sum_{i=1}^{B}\sum_{n=1}^{N} \mathrm{E}\left\{\max_k \log\left(1+P_{\mathsf{con}}\gamma_{i,k}^n\right)\right\} \quad \text{(I.7)}$$

$$= f_{\mathsf{lo}}^{\mathsf{DN}}(r,B,N) \sum_{i,n} \mathrm{E}\left\{\log\left(1+P_{\mathsf{con}} \max_k \gamma_{i,k}^n\right)\right\}, \quad \text{(I.8)}$$

where (I.7) follows because $\sum_{j\neq i} |\nu_{j,k}^n|^2$ is independent of $\nu_{i,k}^n$ (and hence, independent of $\gamma_{i,k}^n$). Note that for Rayleigh fading channels, $\mu = \sigma = 1$. ∎

Lemma I.1 and earlier proved Theorem 1 show that the lower and upper bounds on $\mathcal{C}^*$ are functions of $\max_k \gamma_{i,k}^n$. To compute $\max_k \gamma_{i,k}^n$ for large $K$, we prove two lemmas, Lemma 2 and Lemma 3.

**Lemma I.2.** *Under Rayleigh fading, i.e., $\nu_{i,k}^n \sim \mathcal{CN}(0,1)$, the CDF of $\gamma_{i,k}^n$ is given by*

$$F_{\gamma_{i,k}^n}(\gamma) = 1 - \frac{r_0^2}{p^2}e^{-\frac{\gamma}{\beta^2 r_0^{-2\alpha}}} - \frac{1}{\alpha\beta^2 p^2}\int_{\frac{\beta^2}{(p-d)^{2\alpha}}}^{\beta^2 r_0^{-2\alpha}} e^{-\frac{\gamma}{g}}\left(\frac{g}{\beta^2}\right)^{-1-\frac{1}{\alpha}}dg + \int_{\frac{\beta^2}{(p+d)^{2\alpha}}}^{\frac{\beta^2}{(p-d)^{2\alpha}}} \exp(-\gamma/g)ds(g), \quad \text{(I.9)}$$

where $d = \sqrt{a_i^2+b_i^2}$, and

$$s(g) = \frac{1}{\pi p^2}\left[\left(\frac{g}{\beta^2}\right)^{-1/\alpha}\cos^{-1}\left(\frac{d^2+\left(\frac{g}{\beta^2}\right)^{-1/\alpha}-p^2}{2d\left(\frac{g}{\beta^2}\right)^{-1/2\alpha}}\right) + p^2\cos^{-1}\left(\frac{d^2+p^2-\left(\frac{g}{\beta^2}\right)^{-1/\alpha}}{2dp}\right)\right.$$

$$\left. -\frac{1}{2}\sqrt{\left(p+d-\left(\frac{g}{\beta^2}\right)^{-1/2\alpha}\right)\left(p+\left(\frac{g}{\beta^2}\right)^{-1/2\alpha}-d\right)\left(d+\left(\frac{g}{\beta^2}\right)^{-1/2\alpha}-p\right)\left(d+p+\left(\frac{g}{\beta^2}\right)^{-1/2\alpha}\right)}\right].$$

*Proof:* We assume that the users are distributed uniformly in a circular area of radius $p$ and there are $B$ base-stations in that area as shown in Fig. I.1.

The probability density function of the user-coordinates $(x_k, y_k)$ can be written as

$$f_{(x_k,y_k)}(x,y) = \begin{cases} \frac{1}{\pi p^2} & x^2+y^2 \leq p^2 \\ 0 & \text{otherwise.} \end{cases} \quad \text{(I.10)}$$

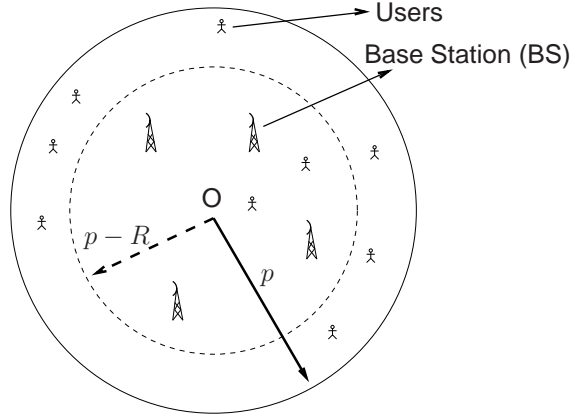

Fig. I.1: OFDMA downlink system with $K$ users and $B$ base-stations.

Note that around any base-station, the users are distributed at-least within a distance $R$ ($R > r_0$). Hence, $p - d = p - \sqrt{a_i^2 + b_i^2} \geq R > r_0$ for all $i$. Now,

$$\gamma_{i,k}^n = \Big(\underbrace{\max\Big\{r_0, \sqrt{(x_k - a_i)^2 + (y_k - b_i)^2}\Big\}}_{R_{i,k}}\Big)^{\overbrace{-2\alpha}^{G_{i,k}}} \beta^2 |\nu_{i,k}^n|^2 \qquad (I.11)$$

$$= \min\Big\{r_0^{-2\alpha}, \big((x_k - a_i)^2 + (y_k - b_i)^2\big)^{-\alpha}\Big\} \beta^2 |\nu_{i,k}^n|^2. \qquad (I.12)$$

We now compute the probability density function of $G_{i,k}$ ($= \beta^2 R_{i,k}^{-2\alpha}$).

$$\begin{aligned} Pr(G_{i,k} > g) &= Pr\Big(r_0^{-2\alpha} > \frac{g}{\beta^2}\Big) \times Pr\Big(\big((x_k - a_i)^2 + (y_k - b_i)^2\big)^{-\alpha} > \frac{g}{\beta^2}\Big) \\ &= Pr\Big(r_0 < \Big(\frac{g}{\beta^2}\Big)^{-1/2\alpha}\Big) \times Pr\Big(\sqrt{(x_k - a_i)^2 + (y_k - b_i)^2} < \Big(\frac{g}{\beta^2}\Big)^{-1/2\alpha}\Big) \\ &= \begin{cases} 0 & \text{if } g \geq \beta^2 r_0^{-2\alpha} \\ Pr\Big(\sqrt{(x_k - a_i)^2 + (y_k - b_i)^2} < \big(\frac{g}{\beta^2}\big)^{-1/2\alpha}\Big) & \text{otherwise.} \end{cases} \end{aligned} \qquad (I.13)$$

Now, $Pr\Big(\sqrt{(x_k - a_i)^2 + (y_k - b_i)^2} < \big(\frac{g}{\beta^2}\big)^{-1/2\alpha}\Big)$ is basically the probability that the distance between the user $k$ and BS $i$ is less than $\big(\frac{g}{\beta^2}\big)^{-1/2\alpha}$. Since, the users are uniformly distributed, this probability is precisely equal to $\frac{1}{\pi p^2}$ times the intersection area of the overall area (of radius $p$ around O) and a circle around BS $i$ with a radius of $\big(\frac{g}{\beta^2}\big)^{-1/2\alpha}$. This is shown as the shaded region in Fig. I.2.

Therefore, we have:

$$Pr(G_{i,k} > g) = \begin{cases} 1 & \text{if } \big(\frac{g}{\beta^2}\big)^{-1/2\alpha} \in (p + d, \infty) \\ s(g) & \text{if } \big(\frac{g}{\beta^2}\big)^{-1/2\alpha} \in (p - d, p + d] \\ \big(\frac{g}{\beta^2}\big)^{-1/\alpha} \frac{1}{p^2} & \text{if } \big(\frac{g}{\beta^2}\big)^{-1/2\alpha} \in (r_0, p - d] \\ 0 & \text{if } \big(\frac{g}{\beta^2}\big)^{-1/2\alpha} \in [0, r_0], \end{cases} \qquad (I.14)$$

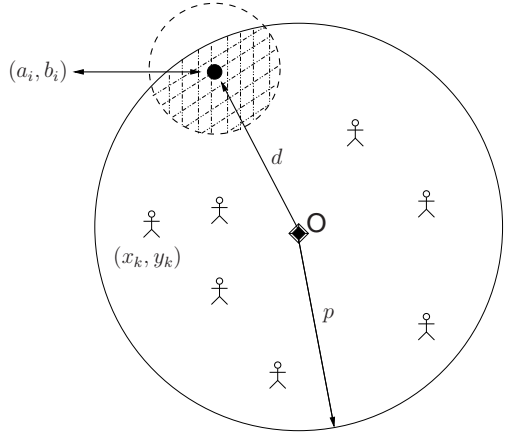

Fig. I.2: System Layout. The BS $i$ is located at a distance of $d$ from the center with the coordinates $(a_i, b_i)$, and the user is stationed at $(x_k, y_k)$.

where

$$s(g) = \frac{1}{\pi p^2}\left[\left(\frac{g}{\beta^2}\right)^{-1/\alpha}\cos^{-1}\left(\frac{d^2 + \left(\frac{g}{\beta^2}\right)^{-1/\alpha} - p^2}{2d\left(\frac{g}{\beta^2}\right)^{-1/2\alpha}}\right) + p^2\cos^{-1}\left(\frac{d^2 + p^2 - \left(\frac{g}{\beta^2}\right)^{-1/\alpha}}{2dp}\right)\right. \quad (\text{I.15})$$

$$\left. -\frac{1}{2}\sqrt{\left(p+d-\left(\frac{g}{\beta^2}\right)^{-1/2\alpha}\right)\left(p+\left(\frac{g}{\beta^2}\right)^{-1/2\alpha}-d\right)\left(d+\left(\frac{g}{\beta^2}\right)^{-1/2\alpha}-p\right)\left(d+p+\left(\frac{g}{\beta^2}\right)^{-1/2\alpha}\right)}\right].$$

The CDF of $G_{i,k}$ can now be written as

$$F_{G_{i,k}}(g) = \begin{cases} 0 & \text{if } g \in \left[0, \beta^2(p+d)^{-2\alpha}\right) \\ 1 - s(g) & \text{if } g \in \left[\beta^2(p+d)^{-2\alpha}, \beta^2(p-d)^{-2\alpha}\right) \\ 1 - \left(\frac{g}{\beta^2}\right)^{-1/\alpha}\frac{1}{p^2} & \text{if } g \in \left[\beta^2(p-d)^{-2\alpha}, \beta^2 r_0^{-2\alpha}\right) \\ 1 & \text{if } g \in \left[\beta^2 r_0^{-2\alpha}, \infty\right), \end{cases} \quad (\text{I.16})$$

A plot of the above CDF is shown in Fig. I.3.

The probability density function of $G_{i,k}$ can be written as follows:

$$f_{G_{i,k}}(g) = \begin{cases} 0 & \text{if } g \in \left[0, \beta^2(p+d)^{-2\alpha}\right) \\ -\frac{ds(g)}{dg} & \text{if } g \in \left[\beta^2(p+d)^{-2\alpha}, \beta^2(p-d)^{-2\alpha}\right) \\ \frac{1}{\alpha\beta^2 p^2}\left(\frac{g}{\beta^2}\right)^{-1-1/\alpha} & \text{if } g \in \left[\beta^2(p-d)^{-2\alpha}, \beta^2 r_0^{-2\alpha}\right) \\ \frac{r_0^2}{p^2} & \text{if } g = \beta^2 r_0^{-2\alpha} \\ 0 & \text{if } g > \beta^2 r_0^{-2\alpha}, \end{cases} \quad (\text{I.17})$$

where $\frac{ds(g)}{dg} \leq 0$. The pdf of $G_{i,k}$ has a discontinuity of the *first-kind* at $\beta^2 r_0^{-2\alpha}$ (where it takes an impulse value), and is continuous in $[\beta^2(p+d)^{-2\alpha}, \beta^2 r_0^{-2\alpha})$. At all other points, it takes the value 0.

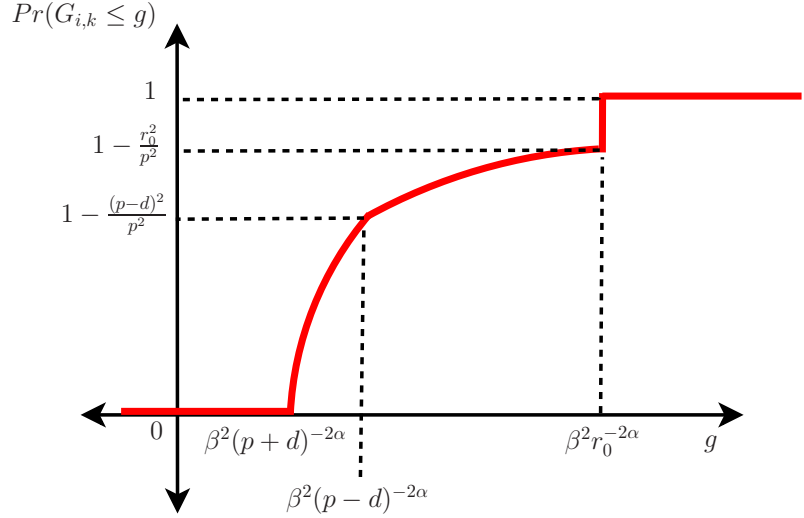

Fig. I.3: Cumulative distribution function of $G_{i,k}$.

Using (I.17), the cumulative distribution function of $\gamma_{i,k}^n$, i.e., $F_{\gamma_{i,k}^n}(\gamma)$ (when $\gamma \geq 0$) can be written as

$$F_{\gamma_{i,k}^n}(\gamma)$$
$$= \int p\Big(|\nu_{i,k}^n|^2 \leq \frac{\gamma}{g}\Big) f_{G_{i,k}}(g) dg \tag{I.18}$$
$$= \int \left(1 - e^{-\gamma/g}\right) f_{G_{i,k}}(g) dg \tag{I.19}$$
$$= 1 - \int e^{-\gamma/g} f_{G_{i,k}}(g) dg \tag{I.20}$$
$$= 1 - \frac{r_0^2}{p^2} e^{-\frac{\gamma}{\beta^2 r_0^{-2\alpha}}} - \int_{\beta^2(p-d)^{-2\alpha}}^{\beta^2 r_0^{-2\alpha}} e^{-\gamma/g} \frac{1}{\alpha \beta^2 p^2} \Big(\frac{g}{\beta^2}\Big)^{-1-1/\alpha} dg + \int_{\beta^2(p+d)^{-2\alpha}}^{\beta^2(p-d)^{-2\alpha}} e^{-\gamma/g} ds(g). \tag{I.21}$$

∎

**Lemma I.3.** *Let $\gamma_{i,k}^n$ be a random variable with a cdf defined in Lemma I.2. Then, the growth function $\frac{1-F_{\gamma_{i,k}^n}(\gamma)}{f_{\gamma_{i,k}^n}(\gamma)}$ converges to a constant $\beta^2 r_0^{-2\alpha}$ as $\gamma \to \infty$, and $\gamma_{i,k}^n$ belongs to a domain of attraction [DN03]. Furthermore, the cdf of $(\max_k \gamma_{i,k}^n - l_K)$ converges in distribution to a limiting random variable with a Gumbel type cdf, that is given by*

$$\exp(-e^{-x r_0^{2\alpha}/\beta^2}), \ x \in (-\infty, \infty), \tag{I.22}$$

*where $l_K$ is such that $F_{\gamma_{i,k}^n}(l_K) = 1 - 1/K$. In particular, $l_K = \beta^2 r_0^{-2\alpha} \log \frac{K r_0^2}{p^2}$.*

*Proof:* We have from Lemma I.2

$$\begin{aligned}
F_{\gamma_{i,k}^n}(\gamma) &= 1 - \frac{r_0^2}{p^2}e^{-\frac{\gamma}{\beta^2 r_0^{-2\alpha}}} - \int_{\beta^2(p-d)^{-2\alpha}}^{\beta^2 r_0^{-2\alpha}} e^{-\gamma/g}\frac{1}{\alpha\beta^2 p^2}\left(\frac{g}{\beta^2}\right)^{-1-1/\alpha}dg + \int_{\beta^2(p+d)^{-2\alpha}}^{\beta^2(p-d)^{-2\alpha}} e^{-\gamma/g}s'(g)dg \\
&= 1 - \frac{r_0^2}{p^2}e^{-\frac{\gamma}{\beta^2 r_0^{-2\alpha}}} - \int_{\beta^2(p-d)^{-2\alpha}}^{\beta^2 r_0^{-2\alpha}} e^{-\gamma/g}\frac{1}{\alpha\beta^2 p^2}\left(\frac{g}{\beta^2}\right)^{-1-1/\alpha}dg + s(g)e^{-\gamma/g}\Big|_{\beta^2(p+d)^{-2\alpha}}^{\beta^2(p-d)^{-2\alpha}} \\
&\quad - \gamma\int_{\beta^2(p+d)^{-2\alpha}}^{\beta^2(p-d)^{-2\alpha}} \frac{e^{-\gamma/g}s(g)}{g^2}dg \quad (\text{I.23}) \\
&= 1 - \frac{r_0^2}{p^2}e^{-\frac{\gamma}{\beta^2 r_0^{-2\alpha}}} - \int_{\beta^2(p-d)^{-2\alpha}}^{\beta^2 r_0^{-2\alpha}} e^{-\frac{\gamma}{g}}\frac{1}{\alpha\beta^2 p^2}\left(\frac{g}{\beta^2}\right)^{-1-\frac{1}{\alpha}}dg + e^{-\frac{\gamma}{\beta^2(p-d)^{-2\alpha}}}\frac{(p-d)^2}{p^2} \\
&\quad - e^{-\frac{\gamma}{\beta^2(p+d)^{-2\alpha}}} - \gamma\int_{\beta^2(p+d)^{-2\alpha}}^{\beta^2(p-d)^{-2\alpha}} \frac{e^{-\frac{\gamma}{g}}s(g)}{g^2}dg, \quad (\text{I.24})
\end{aligned}$$

where $\frac{r_0^2}{p^2} < \frac{(p-d)^2}{p^2} \leq s(g) \leq 1$ (see Fig. I.3). Now, we claim that

$$\lim_{\gamma\to\infty}\left(1 - F_{\gamma_{i,k}^n}(\gamma)\right)e^{\frac{\gamma}{\beta^2 r_0^{-2\alpha}}} = \frac{r_0^2}{p^2}. \quad (\text{I.25})$$

It is clear that the first two terms in (I.24) contribute everything to the limit in (I.25). We will consider the rest of the terms now and show that they contribute zero towards the limit in RHS of (I.25). First, considering the 4$^{\text{th}}$, 5$^{\text{th}}$, and 6$^{\text{th}}$ terms, we have

$$\lim_{\gamma\to\infty} e^{\frac{\gamma}{\beta^2 r_0^{-2\alpha}}} \times \left| e^{-\frac{\gamma}{\beta^2(p-d)^{-2\alpha}}}\frac{(p-d)^2}{p^2} - e^{-\frac{\gamma}{\beta^2(p+d)^{-2\alpha}}} - \gamma\int_{\beta^2(p+d)^{-2\alpha}}^{\beta^2(p-d)^{-2\alpha}} \frac{e^{-\frac{\gamma}{g}}s(g)}{g^2}dg \right| \quad (\text{I.26})$$

$$\leq \lim_{\gamma\to\infty} e^{\frac{\gamma}{\beta^2 r_0^{-2\alpha}}}\left(\left|e^{-\frac{\gamma}{\beta^2(p-d)^{-2\alpha}}}\frac{(p-d)^2}{p^2}\right| + \left|e^{-\frac{\gamma}{\beta^2(p+d)^{-2\alpha}}}\right| + \left|\gamma\int_{\beta^2(p+d)^{-2\alpha}}^{\beta^2(p-d)^{-2\alpha}} \frac{e^{-\frac{\gamma}{g}}s(g)}{g^2}dg\right|\right) \quad (\text{I.27})$$

$$\leq \lim_{\gamma\to\infty} \frac{(p-d)^2}{p^2}e^{-\frac{\gamma}{\beta^2}((p-d)^{2\alpha}-r_0^{2\alpha})} + e^{-\frac{\gamma}{\beta^2}((p+d)^{2\alpha}-r_0^{2\alpha})} + \gamma\frac{e^{-\frac{\gamma}{\beta^2}((p-d)^{2\alpha}-r_0^{2\alpha})}}{\beta^4(p+d)^{-4\alpha}} \quad (\text{I.28})$$

$$= 0. \quad (\text{I.29})$$

Now, we consider the third term in (I.24). We will show that

$$\lim_{\gamma\to\infty} \underbrace{e^{\frac{\gamma}{\beta^2 r_0^{-2\alpha}}} \times \int_{\beta^2(p-d)^{-2\alpha}}^{\beta^2 r_0^{-2\alpha}} e^{-\frac{\gamma}{g}}\frac{1}{\alpha\beta^2 p^2}\left(\frac{g}{\beta^2}\right)^{-1-\frac{1}{\alpha}}dg}_{\mathcal{T}(\gamma)} = 0. \quad (\text{I.30})$$

Taking the exponential inside the integral, we have

$$\mathcal{T}(\gamma) = \int_{\beta^2(p-d)^{-2\alpha}}^{\beta^2 r_0^{-2\alpha}} e^{-\frac{\gamma}{g}+\gamma r_0^{2\alpha}/\beta^2}\frac{1}{\alpha\beta^2 p^2}\left(\frac{g}{\beta^2}\right)^{-1-\frac{1}{\alpha}}dg. \quad (\text{I.31})$$

Substituting $\gamma/g$ by $x$, we get

$$\mathcal{T}(\gamma) = \int_{\frac{\gamma r_0^{2\alpha}}{\beta^2}}^{\frac{\gamma(p-d)^{2\alpha}}{\beta^2}} e^{-x+\gamma r_0^{2\alpha}/\beta^2}\frac{1}{\alpha\beta^2 p^2}\left(\frac{\gamma}{x\beta^2}\right)^{-1-\frac{1}{\alpha}}\left(\frac{\gamma}{x^2}\right)dx. \quad (\text{I.32})$$

Again substituting $x - \gamma r_0^{2\alpha}/\beta^2$ by $y$, we have

$$\mathcal{T}(\gamma) = \frac{1}{\alpha p^2}\left(\frac{\gamma}{\beta^2}\right)^{-\frac{1}{\alpha}} \int_0^{\gamma\frac{(p-d)^{2\alpha}-r_0^{2\alpha}}{\beta^2}} e^{-y}\left(y + \frac{\gamma r_0^{2\alpha}}{\beta^2}\right)^{-1+\frac{1}{\alpha}} dy \tag{I.33}$$

$$\leq \frac{1}{\alpha p^2}\left(\frac{\gamma}{\beta^2}\right)^{-\frac{1}{\alpha}} \left(\frac{\gamma r_0^{2\alpha}}{\beta^2}\right)^{-1+\frac{1}{\alpha}} \int_0^{\frac{\left((p-d)^{2\alpha}-r_0^{2\alpha}\right)\gamma}{\beta^2}} e^{-y} dy \tag{I.34}$$

$$= \frac{\beta^2}{\alpha\gamma p^2} r_0^{-2\alpha+2}\left(1 - e^{-\gamma\frac{(p-d)^{2\alpha}-r_0^{2\alpha}}{\beta^2}}\right), \tag{I.35}$$

where, in (I.34), an upper bound is taken by putting $y = 0$ in the term $\left(y + \frac{\gamma r_0^{2\alpha}}{\beta^2}\right)^{-1+\frac{1}{\alpha}}$ inside the integral. Since $\mathcal{T}(\gamma)$ is positive, (I.35) shows that $\lim_{\gamma \to \infty} \mathcal{T}(\gamma) = 0$. Hence, the claim is true.

Now, after computing the derivative of $F_{\gamma_{i,k}^n}(\gamma)$ w.r.t. $\gamma$ to obtain the probability density function $f_{\gamma_{i,k}^n}(\gamma)$, we have

$$\lim_{\gamma \to \infty} f_{\gamma_{i,k}^n}(\gamma) e^{\gamma r_0^{2\alpha}/\beta^2} = \frac{r_0^2}{p^2 \beta^2 r_0^{-2\alpha}}. \tag{I.36}$$

We do not prove the above equation here as (I.36) is straightforward to verify (similar to the steps taken to prove (I.25)). From (I.25) and (I.36), we obtain that the growth function converges to a constant, i.e.,

$$\lim_{\gamma \to \infty} \frac{1 - F_{\gamma_{i,k}^n}(\gamma)}{f_{\gamma_{i,k}^n}(\gamma)} = \beta^2 r_0^{-2\alpha}. \tag{I.37}$$

The above equation implies that $\gamma_{i,k}^n$ belongs to a *domain of maximal attraction* [DN03, pp. 296]. In particular, the cdf of $(\max_k \gamma_{i,k}^n - l_K)$ converges in distribution to a limiting random variable with an extreme-value cdf, that is given by

$$\exp(-e^{-xr_0^{2\alpha}/\beta^2}), \; x \in (-\infty, \infty). \tag{I.38}$$

Here, $l_K$ is such that $F_{\gamma_{i,\cdot}^n}(l_K) = 1 - 1/K$. Solving for $l_K$, we have

$$\frac{1}{K} = \frac{r_0^2}{p^2} e^{-\frac{l_K}{\beta^2 r_0^{-2\alpha}}} + \int_{\frac{\beta^2}{(p-d)^{2\alpha}}}^{\frac{\beta^2}{r_0^{2\alpha}}} e^{-\frac{l_K}{g}} \frac{1}{\alpha\beta^2 p^2}\left(\frac{g}{\beta^2}\right)^{-1-\frac{1}{\alpha}} dg + \int_{\frac{\beta^2}{(p+d)^{2\alpha}}}^{\frac{\beta^2}{(p-d)^{2\alpha}}} e^{-\frac{l_K}{g}}\left(-s'(g)\right) dg \tag{I.39}$$

Substituting $l_K/g$ by $x$ in the first integral in RHS of (I.39) and computing an upper bound, we get

$$\frac{1}{K} \leq \frac{r_0^2}{p^2} e^{-\frac{l_K}{\beta^2 r_0^{-2\alpha}}} + \frac{1}{\alpha\beta^2 p^2} \int_{\frac{l_K}{\beta^2(p-d)^{-2\alpha}}}^{\frac{l_K}{\beta^2 r_0^{-2\alpha}}} e^{-x} \left(\frac{l_K}{x\beta^2}\right)^{-1-\frac{1}{\alpha}} \left(\frac{-l_K}{x^2}\right) dx - e^{-\frac{l_K}{\beta^2(p-d)^{-2\alpha}}} \int_{\frac{\beta^2}{(p+d)^{2\alpha}}}^{\frac{\beta^2}{(p-d)^{2\alpha}}} s'(g) dg \quad \text{(I.40)}$$

$$= \exp\left(-\frac{l_K}{\beta^2 r_0^{-2\alpha}}\right) \frac{r_0^2}{p^2} + \frac{1}{\alpha p^2} \left(\frac{l_K}{\beta^2}\right)^{-\frac{1}{\alpha}} \int_{\frac{l_K}{\beta^2 r_0^{-2\alpha}}}^{\frac{l_K}{\beta^2(p-d)^{-2\alpha}}} e^{-x} x^{-1+\frac{1}{\alpha}} dx + e^{-\frac{l_K}{\beta^2(p-d)^{-2\alpha}}} \left(-s(g)\right)\Big|_{g=\frac{\beta^2}{(p+d)^{2\alpha}}}^{g=\frac{\beta^2}{(p-d)^{2\alpha}}}$$

$$\leq \frac{r_0^2}{p^2} e^{-\frac{l_K}{\beta^2 r_0^{-2\alpha}}} + \frac{1}{\alpha p^2} \left(\frac{l_K}{\beta^2}\right)^{-\frac{1}{\alpha}} \left(\frac{l_K r_0^{2\alpha}}{\beta^2}\right)^{-1+\frac{1}{\alpha}} \int_{\frac{l_K}{\beta^2 r_0^{-2\alpha}}}^{\frac{l_K}{\beta^2(p-d)^{-2\alpha}}} e^{-x} dx + e^{-\frac{l_K}{\beta^2(p-d)^{-2\alpha}}} \left(1 - \frac{(p-d)^2}{p^2}\right) \quad \text{(I.41)}$$

$$\leq e^{-\frac{l_K}{\beta^2 r_0^{-2\alpha}}} \frac{r_0^2}{p^2} + \frac{r_0^{2-2\alpha}}{\alpha p^2} \left(\frac{l_K}{\beta^2}\right)^{-1} \int_{\frac{l_K}{\beta^2 r_0^{-2\alpha}}}^{\infty} e^{-x} dx + e^{-\frac{l_K}{\beta^2(p-d)^{-2\alpha}}} \quad \text{(I.42)}$$

$$\leq e^{-\frac{l_K}{\beta^2 r_0^{-2\alpha}}} \frac{r_0^2}{p^2} + \frac{r_0^{2-2\alpha}}{\alpha p^2} \left(\frac{l_K}{\beta^2}\right)^{-1} e^{-\frac{l_K}{\beta^2 r_0^{-2\alpha}}} + e^{-\frac{l_K}{\beta^2(p-d)^{-2\alpha}}} \quad \text{(I.43)}$$

$$\leq e^{-\frac{l_K}{\beta^2 r_0^{-2\alpha}}} \frac{r_0^2}{p^2} \left(1 + \frac{\beta^2 r_0^{-2\alpha}}{\alpha l_K} + \frac{p^2}{r_0^2} e^{-\frac{l_K}{\beta^2}\left((p-d)^{2\alpha} - r_0^{2\alpha}\right)}\right) \quad \text{(I.44)}$$

$$= e^{-\frac{l_K}{\beta^2 r_0^{-2\alpha}}} \frac{r_0^2}{p^2} \left(1 + O\left(\frac{1}{l_K}\right)\right). \quad \text{(I.45)}$$

In (I.40), we substitute $l_K/g$ by $x$ in the first integral of (I.39), and compute an upper bound by taking the exponential term out of the second integral of (I.39). In (I.41), we note that $\frac{(p-d)^2}{p^2} \leq s(g) \leq 1$. From the above analysis, we now have

$$l_K \leq \beta^2 r_0^{-2\alpha} \log \frac{K r_0^2}{p^2} + O\left(\frac{1}{l_K}\right). \quad \text{(I.46)}$$

Now, to compute a lower bound on $l_K$ from (I.39), we note that fact that $\frac{ds(g)}{dg} \leq 0$. Therefore,

$$\frac{1}{K} \geq \frac{r_0^2}{p^2} e^{-\frac{l_K}{\beta^2 r_0^{-2\alpha}}} \quad \text{(I.47)}$$

$$\implies l_K \geq \beta^2 r_0^{-2\alpha} \log \frac{K r_0^2}{p^2}. \quad \text{(I.48)}$$

From (I.46) and (I.48), we have $\beta^2 r_0^{-2\alpha} \log \frac{K r_0^2}{p^2} \leq l_K \leq \beta^2 r_0^{-2\alpha} \log \frac{K r_0^2}{p^2} + O\left(\frac{1}{\log K}\right)$. Therefore,

$$l_K \approx \beta^2 r_0^{-2\alpha} \log \frac{K r_0^2}{p^2} \quad \text{(I.49)}$$

for large $K$. ∎

Interestingly, for a given BS $i$, the scaling of $\max_k \gamma_{i,k}^n$ (given by $l_K$ in large $K$ regime) is independent of the coordinates $(a_i, b_i)$ and is a function of $r_0, p$. Now, since the growth function converges to a constant (Lemma I.3), we apply [SH05, Theorem A.2] giving us:

$$Pr\left\{l_K - \log\log K \leq \max_k \gamma_{i,k}^n \leq l_K + \log\log K\right\} \geq 1 - O\left(\frac{1}{\log K}\right), \quad \text{(I.50)}$$

where $l_K = \beta^2 r_0^{-2\alpha} \log \frac{K r_0^2}{p^2}$. Therefore,

$$\begin{aligned}
\mathrm{E}\left\{\log\left(1 + P_{\mathsf{con}} \max_k \gamma_{i,k}^n\right)\right\} &\leq Pr\left(\max_k \gamma_{i,k}^n \leq l_K + \log\log K\right) \log(1 + P_{\mathsf{con}} l_K + P_{\mathsf{con}} \log\log K) \\
&\quad + Pr\left(\max_k \gamma_{i,k}^n > l_K + \log\log K\right) \log(1 + P_{\mathsf{con}} \beta^2 r_0^{-2\alpha} K) \quad (\text{I.51})\\
&\leq \log(1 + P_{\mathsf{con}} l_K + P_{\mathsf{con}} \log\log K) + \log(1 + P_{\mathsf{con}} \beta^2 r_0^{-2\alpha} K) \times O\left(\frac{1}{\log K}\right) \\
&= \log(1 + P_{\mathsf{con}} l_K) + O(1). \quad (\text{I.52})
\end{aligned}$$

where, in (I.51), we have used the fact that the sum-rate is bounded above by $\log(1 + P_{\mathsf{con}} \beta^2 r_0^{-2\alpha} K)$. This is because

$$\begin{aligned}
\mathrm{E}\left\{\log\left(1 + P_{\mathsf{con}} \max_k \gamma_{i,k}^n\right)\right\} &\leq \mathrm{E}\left\{\log\left(1 + P_{\mathsf{con}} \sum_k \gamma_{i,k}^n\right)\right\} \\
&\leq \log\left(1 + P_{\mathsf{con}} \sum_k \mathrm{E}\{\gamma_{i,k}^n\}\right) \quad (\text{I.53})\\
&\leq \log\left(1 + P_{\mathsf{con}} \beta^2 r_0^{-2\alpha} \sum_k \mathrm{E}\{|\nu_{i,k}^n|^2\}\right) \\
&\leq \log\left(1 + P_{\mathsf{con}} \beta^2 r_0^{-2\alpha} K\right), \quad (\text{I.54})
\end{aligned}$$

where (I.53) follows from Jensen's inequality. Further, from (I.50), we have

$$\mathrm{E}\left\{\log\left(1 + P_{\mathsf{con}} \max_k \gamma_{i,k}^n\right)\right\} \geq \log(1 + P_{\mathsf{con}} l_K - P_{\mathsf{con}} \log\log K)\left(1 - O\left(\frac{1}{\log K}\right)\right). \quad (\text{I.55})$$

Combining (I.52) and (I.55), we get, for large $K$,

$$\begin{aligned}
BN \log(1 + P_{\mathsf{con}} l_K - P_{\mathsf{con}} \log\log K)\left(1 - O\left(\frac{1}{\log K}\right)\right) &\leq \sum_{i,n} \mathrm{E}\left\{\log\left(1 + P_{\mathsf{con}} \max_k \gamma_{i,k}^n\right)\right\} \quad (\text{I.56})\\
&\leq \left(\log(1 + P_{\mathsf{con}} l_K) + O(1)\right) BN.
\end{aligned}$$

Therefore, from Lemma I.1 and Theorem 1, we get

$$\left(\log(1 + P_{\mathsf{con}} l_K) + O(1)\right) BN f_{\mathsf{lo}}^{\mathsf{DN}}(r, B, N) \leq \mathcal{C}^* \leq \left(\log(1 + P_{\mathsf{con}} l_K) + O(1)\right) BN \quad (\text{I.57})$$

This results in:

$$\begin{aligned}
\mathcal{C}^* &= O(BN \log\log K), \text{ and} \\
\mathcal{C}^* &= \Omega(BN f_{\mathsf{lo}}^{\mathsf{DN}}(r, B, N) \log\log K). \quad (\text{I.58})
\end{aligned}$$

Now, to prove Corollary 1, we use the upper bound in Theorem 1 obtained via Jensen's inequality. In particular, we have

$$\begin{aligned}
\mathcal{C}^* &\leq N \sum_i \mathrm{E}\left\{\log\left(1 + \frac{P_{\mathsf{con}}}{N} \max_{n,k} \gamma_{i,k}^n\right)\right\} \quad (\text{I.59})\\
&\leq BN \log\left(1 + \frac{P_{\mathsf{con}}}{N} l_{KN}\right) + BN\, O(1), \quad (\text{I.60})
\end{aligned}$$

where (I.60) follows from (I.56), and $l_{KN} = \beta^2 r_0^{-2\alpha} \log \frac{KNr_0^2}{p^2}$ determines the SNR scaling of the maximum over $KN$ i.i.d. random variables. This implies

$$\mathcal{C}^* = O\Big(BN \log \frac{\log KN}{N}\Big). \tag{I.61}$$

Note that the above result is only true if $\frac{P_{\text{con}}}{N} l_{KN} \gg 1$, (or, $\log \frac{KN}{N} \gg 1$ to make the approximation $\log(1+x) \approx \log x$ valid for large $x$).

## II. PROOF OF THEOREM 3 AND COROLLARY 2 FOR DENSE NETWORKS

We will first find the SNR scaling laws for each of the three families of distributions — Nakagami-$m$, Weibull, and LogNormal. This involves deriving the domain of attraction of channel-SNR $\gamma_{i,k}^n$ for all three types of distributions. The domains of attraction are of three types - Fréchet, Weibull, and Gumbel. Let the growth function be defined as $h(\gamma) \triangleq \frac{1-F_{\gamma_{i,k}^n}(\gamma)}{f_{\gamma_{i,k}^n}(\gamma)}$. The random variable, $\gamma_{i,k}^n$, belongs to the Gumbel-type if $\lim_{\gamma \to \infty} h'(\gamma) = 0$. It turns out that all three distributions considered in [AKS12], i.e., Nakagami-$m$, Weibull, and LogNormal, belong to this category. After showing this, we find the scaling, $l_K$, such that $F_{\gamma_{i,k}^n}(l_K) = 1 - 1/K$. The intuition behind this choice of $l_K$ is that the cdf of $\max_k \gamma_{i,k}^n$ is $F_{\gamma_{i,k}^n}^K(\gamma)$. For $\gamma = l_K$, we have $F_{\gamma_{i,k}^n}^K(l_K) = (1 - 1/K)^K \to e^{-1}$. The fact that $F_{\gamma_{i,k}^n}^K(\gamma)$ converges for a particular choice of $\gamma$ gives information about the asymptotic behavior of $\max_k \gamma_{i,k}^n$.

### A. Nakagami-$m$

In this case, $|\nu_{i,k}^n|$ is distributed according to Nakagami-$(m, w)$ distribution. Hence, $|\nu_{i,k}^n|^2$ is distributed according to Gamma-$(m, w/m)$ distribution. The cumulative distribution function of $\gamma_{i,k}^n$, i.e., $F_{\gamma_{i,k}^n}(\gamma)$ (when $\gamma \geq 0$) is

$$F_{\gamma_{i,k}^n}(\gamma) = \int p\Big(|\nu_{i,k}^n|^2 \leq \frac{\gamma}{g}\Big) f_{G_{i,k}}(g) dg \tag{II.1}$$

$$= \int \frac{\gamma(m, \frac{m\gamma}{wg})}{\Gamma(m)} f_{G_{i,k}}(g) dg \tag{II.2}$$

$$= 1 - \int_{\beta^2(p+d)^{-2\alpha}}^{\beta^2 r_0^{-2\alpha}} \frac{\Gamma(m, \frac{m\gamma}{wg})}{\Gamma(m)} f_{G_{i,k}}(g) dg \tag{II.3}$$

where $f_{G_{i,k}}(g)$ is defined in (I.17). Now, for large $\gamma$, we can approximate (II.3) as

$$F_{\gamma_{i,k}^n}(\gamma) \approx 1 - \frac{1}{\Gamma(m)} \int_{\beta^2(p+d)^{-2\alpha}}^{\beta^2 r_0^{-2\alpha}} \Big(\frac{m\gamma}{wg}\Big)^{m-1} e^{-\frac{m\gamma}{wg}} f_{G_{i,k}}(g) dg \tag{II.4}$$

$$= 1 - \frac{r_0^2}{p^2 \Gamma(m)} \Big(\frac{m\gamma}{w\beta^2 r_0^{-2\alpha}}\Big)^{m-1} e^{-\frac{m\gamma}{w\beta^2 r_0^{-2\alpha}}} - \frac{1}{\Gamma(m)} \int_{\beta^2(p-d)^{-2\alpha}}^{\beta^2 r_0^{-2\alpha}} \Big(\frac{m\gamma}{wg}\Big)^{m-1} e^{-\frac{m\gamma}{wg}} \frac{1}{\alpha \beta^2 p^2} \Big(\frac{g}{\beta^2}\Big)^{-1-\frac{1}{\alpha}} dg$$

$$+ \frac{1}{\Gamma(m)} \int_{\beta^2(p+d)^{-2\alpha}}^{\beta^2(p-d)^{-2\alpha}} \Big(\frac{m\gamma}{wg}\Big)^{m-1} e^{-\frac{m\gamma}{wg}} ds(g), \tag{II.5}$$

where $f_{G_{i,k}}(g)$ is defined in (I.17). We claim that

$$\lim_{\gamma \to \infty} \left(1 - F_{\gamma_{i,k}^n}(\gamma)\right)\gamma^{1-m} e^{\frac{m\gamma}{w\beta^2 r_0^{-2\alpha}}}$$

$$= \lim_{\gamma \to \infty} \gamma^{1-m} e^{\frac{m\gamma}{w\beta^2 r_0^{-2\alpha}}} \frac{1}{\Gamma(m)} \int_{\beta^2(p+d)^{-2\alpha}}^{\beta^2 r_0^{-2\alpha}} \left(\frac{m\gamma}{wg}\right)^{m-1} e^{-\frac{m\gamma}{wg}} f_{G_{i,k}}(g) dg \quad \text{(II.6)}$$

$$= \frac{r_0^2 m^{m-1}}{p^2 \Gamma(m)(w\beta^2 r_0^{-2\alpha})^{m-1}}. \quad \text{(II.7)}$$

Note that the first two terms in the RHS of (II.5) contribute everything towards the limit in (II.7). We will show that the rest of the terms contribute zero to the limit in RHS of (II.7). In particular, ignoring the constant $\Gamma(m)$, the contribution of the two integral-terms (in (II.5)) is

$$\gamma^{1-m} e^{\frac{m\gamma}{w\beta^2 r_0^{-2\alpha}}} \left( -\int_{\beta^2(p-d)^{-2\alpha}}^{\beta^2 r_0^{-2\alpha}} \left(\frac{m\gamma}{wg}\right)^{m-1} e^{-\frac{m\gamma}{wg}} \frac{1}{\alpha\beta^2 p^2} \left(\frac{g}{\beta^2}\right)^{-1-\frac{1}{\alpha}} dg + \int_{\beta^2(p+d)^{-2\alpha}}^{\beta^2(p-d)^{-2\alpha}} \left(\frac{m\gamma}{wg}\right)^{m-1} e^{-\frac{m\gamma}{wg}} ds(g) \right)$$

$$= \underbrace{-\int_{\beta^2(p-d)^{-2\alpha}}^{\beta^2 r_0^{-2\alpha}} \left(\frac{m}{wg}\right)^{m-1} e^{-\frac{m\gamma}{w}\left(\frac{1}{g} - \frac{1}{\beta^2 r_0^{-2\alpha}}\right)} \frac{1}{\alpha\beta^2 p^2} \left(\frac{g}{\beta^2}\right)^{-1-\frac{1}{\alpha}} dg}_{\mathcal{T}_1(\gamma)} + \underbrace{\int_{\beta^2(p+d)^{-2\alpha}}^{\beta^2(p-d)^{-2\alpha}} \left(\frac{m}{wg}\right)^{m-1} e^{-\frac{m\gamma}{w}\left(\frac{1}{g} - \frac{1}{\beta^2 r_0^{-2\alpha}}\right)} ds(g)}_{\mathcal{T}_2(\gamma)}$$

$$= \mathcal{T}_1(\gamma) + \mathcal{T}_2(\gamma). \quad \text{(II.8)}$$

Now,

$$|\mathcal{T}_1(\gamma)|$$

$$= \left(\frac{m}{w}\right)^{m-1} \frac{\beta^{\frac{2}{\alpha}}}{\alpha p^2} \int_{\frac{\beta^2}{(p-d)^{2\alpha}}}^{\frac{\beta^2}{r_0^{2\alpha}}} g^{-m-\frac{1}{\alpha}} e^{-\frac{m\gamma}{w}\left(\frac{1}{g} - \frac{1}{\beta^2 r_0^{-2\alpha}}\right)} dg \quad \text{(II.9)}$$

$$= \left(\frac{m}{w}\right)^{m-1} \frac{\beta^{\frac{2}{\alpha}}}{\alpha p^2} \int_{\beta^{-2} r_0^{2\alpha}}^{\beta^{-2}(p-d)^{2\alpha}} x^{m+\frac{1}{\alpha}-2} e^{-\frac{m\gamma}{w}\left(x - \beta^{-2} r_0^{2\alpha}\right)} dx \quad \text{(II.10)}$$

$$\leq \left(\frac{m}{w}\right)^{m-1} \frac{\beta^{\frac{2}{\alpha}}}{\alpha p^2} \max\left\{\left(\frac{(p-d)^{2\alpha}}{\beta^2}\right)^{m+\frac{1}{\alpha}-2}, \left(\frac{r_0^{2\alpha}}{\beta^2}\right)^{m+\frac{1}{\alpha}-2}\right\} \int_{\beta^{-2} r_0^{2\alpha}}^{\beta^{-2}(p-d)^{2\alpha}} e^{-\frac{m\gamma}{w}\left(x - \beta^{-2} r_0^{2\alpha}\right)} dx$$

$$= \left(\frac{m}{w}\right)^{m-1} \frac{\beta^{\frac{2}{\alpha}}}{\alpha p^2} \max\left\{\left(\frac{(p-d)^{2\alpha}}{\beta^2}\right)^{m+\frac{1}{\alpha}-2}, \left(\frac{r_0^{2\alpha}}{\beta^2}\right)^{m+\frac{1}{\alpha}-2}\right\} \frac{1 - e^{-\frac{m\gamma}{w}\left(\beta^{-2}(p-d)^{2\alpha} - \beta^{-2} r_0^{2\alpha}\right)}}{\frac{m\gamma}{w}} \quad \text{(II.11)}$$

$$\to 0, \text{ as } \gamma \to \infty. \quad \text{(II.12)}$$

where, in (II.10), we substituted $\frac{1}{g}$ by $x$. Further,

$$|\mathcal{T}_2(\gamma)| = \left| \int_{\beta^2(p+d)^{-2\alpha}}^{\beta^2(p-d)^{-2\alpha}} \left(\frac{m}{wg}\right)^{m-1} e^{-\frac{m\gamma}{w}\left(\frac{1}{g} - \frac{1}{\beta^2 r_0^{-2\alpha}}\right)} ds(g) \right| \quad \text{(II.13)}$$

$$\leq e^{-\frac{m\gamma}{w\beta^2}\left((p-d)^{2\alpha} - r_0^{2\alpha}\right)} \left| \int_{\beta^2(p+d)^{-2\alpha}}^{\beta^2(p-d)^{-2\alpha}} \left(\frac{m}{wg}\right)^{m-1} ds(g) \right| \quad \text{(II.14)}$$

$$\to 0, \text{ as } \gamma \to \infty. \quad \text{(II.15)}$$

Therefore, $\mathcal{T}_1(\gamma)$ and $\mathcal{T}_2(\gamma)$ have zero contribution to the RHS in (II.7), and the our claim is true. Now, from (II.5), we have

$$f_{\gamma_{i,k}^n}(\gamma) \tag{II.16}$$
$$= \frac{\gamma^{m-1}}{\Gamma(m)} \int_{\beta^2(p+d)^{-2\alpha}}^{\beta^2 r_0^{-2\alpha}} \left(\frac{m}{wg}\right)^m e^{-\frac{m\gamma}{wg}} f_{G_{i,k}}(g) dg - \frac{(m-1)\gamma^{m-2}}{\Gamma(m)} \int_{\beta^2(p+d)^{-2\alpha}}^{\beta^2 r_0^{-2\alpha}} \left(\frac{m}{wg}\right)^{m-1} e^{-\frac{m\gamma}{wg}} f_{G_{i,k}}(g) dg$$

Using (II.6)-(II.7), it is easy to verify that

$$\lim_{\gamma \to \infty} f_{\gamma_{i,k}^n}(\gamma) \gamma^{1-m} e^{\frac{m\gamma}{w\beta^2 r_0^{-2\alpha}}} = \frac{r_0^2 m^m}{p^2 \Gamma(m)(w\beta^2 r_0^{-2\alpha})^m}. \tag{II.17}$$

From (II.7) and (II.17), we obtain that the growth function converges to a constant. In particular,

$$\lim_{\gamma \to \infty} \frac{1 - F_{\gamma_{i,k}^n}(\gamma)}{f_{\gamma_{i,k}^n}(\gamma)} = \frac{w\beta^2 r_0^{-2\alpha}}{m}, \tag{II.18}$$

Hence, $\gamma_{i,k}^n$ belongs to the Gumbel-type [DN03] and $\max_k \gamma_{i,k}^n - l_K$ converges in distribution to a limiting random variable with a Gumbel-type cdf, that is given by

$$\exp(-e^{-xr_0^{2\alpha}/\beta^2}), \ x \in (-\infty, \infty), \tag{II.19}$$

where $1 - F_{\gamma_{i,k}^n}(l_K) = \frac{1}{K}$. From (II.7), we have $l_K \approx \frac{w\beta^2 r_0^{-2\alpha}}{m} \log \frac{K r_0^2 m^{m-1}}{p^2 \Gamma(m)(w\beta^2 r_0^{-2\alpha})^{m-1}}$ for large $K$.

Now, since the growth function converges to a constant and $l_K = \Theta(\log K)$, we can use [SH05, Theorem 1] to obtain:

$$Pr\left\{l_K - \log\log K \leq \max_k \gamma_{i,k}^n \leq l_K + \log\log K\right\} \geq 1 - O\left(\frac{1}{\log K}\right). \tag{II.20}$$

This is the same as (I.50). Thus, following the same analysis as in (I.51)-(I.61), we get

$$\mathcal{C}^* = O(BN \log\log \frac{Kr_0^2}{p^2}) \text{ and} \tag{II.21}$$

$$\mathcal{C}^* = BN f_{\text{lo}}^{\text{DN}}(r, B, N) \Omega\left(\log\log \frac{Kr_0^2}{p^2}\right). \tag{II.22}$$

Further, if $\log \frac{KN}{N} \gg 1$, then $\mathcal{C}^* = O\left(\log \frac{\log \frac{KNr_0^2}{p^2}}{N}\right)$.

## B. Weibull

In this case, $|\nu_{i,k}^n|$ is distributed according to Weibull-$(\lambda, t)$ distribution. Hence, $|\nu_{i,k}^n|^2$ is distributed according to Weibull-$(\lambda^2, t/2)$ distribution. We start with finding the cumulative distribution function of $\gamma_{i,k}^n$, i.e., $F_{\gamma_{i,k}^n}(\gamma)$ (when $\gamma \geq 0$) as

$$F_{\gamma_{i,k}^n}(\gamma)$$
$$= \int p\left(|\nu_{i,k}^n|^2 \leq \frac{\gamma}{g}\right) f_{G_{i,k}}(g) dg \tag{II.23}$$
$$= 1 - \int_{\beta^2(p+d)^{-2\alpha}}^{\beta^2 r_0^{-2\alpha}} e^{-\left(\frac{\gamma}{g\lambda^2}\right)^{t/2}} f_{G_{i,k}}(g) dg \tag{II.24}$$
$$= 1 - \frac{r_0^2}{p^2} e^{-\left(\frac{\gamma}{\beta^2 r_0^{-2\alpha}\lambda^2}\right)^{t/2}} - \int_{\frac{\beta^2}{(p-d)^{2\alpha}}}^{\frac{\beta^2}{r_0^{2\alpha}}} \frac{e^{-\left(\frac{\gamma}{g\lambda^2}\right)^{t/2}}}{\alpha\beta^2 p^2} \left(\frac{g}{\beta^2}\right)^{-1-\frac{1}{\alpha}} dg + \int_{\frac{\beta^2}{(p+d)^{2\alpha}}}^{\frac{\beta^2}{(p-d)^{2\alpha}}} e^{-\left(\frac{\gamma}{g\lambda^2}\right)^{t/2}} ds(g). \tag{II.25}$$

This case is similar to the Rayleigh distribution scenario in (I.21). Therefore, it is easy to verify that

$$\lim_{\gamma \to \infty} \left(1 - F_{\gamma_{i,k}^n}(\gamma)\right) e^{\left(\frac{\gamma}{\beta^2 r_0^{-2\alpha} \lambda^2}\right)^{t/2}} = \frac{r_0^2}{p^2}, \text{ and} \tag{II.26}$$

$$\lim_{\gamma \to \infty} f_{\gamma_{i,k}^n}(\gamma) \gamma^{1-t/2} e^{\left(\frac{\gamma}{\beta^2 r_0^{-2\alpha} \lambda^2}\right)^{t/2}} = \frac{t r_0^2}{2\left(\beta^2 r_0^{-2\alpha} \lambda^2\right)^{t/2} p^2}. \tag{II.27}$$

Thus, the growth function $h(\gamma) = \frac{1 - F_{\gamma_{i,k}^n}(\gamma)}{f_{\gamma_{i,k}^n}(\gamma)}$ can be approximated for large $\gamma$ as

$$h(\gamma) \approx \frac{2\left(\beta^2 r_0^{-2\alpha} \lambda^2\right)^{t/2}}{t} \gamma^{1-t/2}. \tag{II.28}$$

Since $\lim_{\gamma \to \infty} h'(\gamma) = 0$, the limiting distribution of $\max_k \gamma_{i,k}^n$ is of Gumbel-type. Note that this is true even when $t < 1$ which refers to heavy-tail distributions. Solving for $1 - F_{\gamma_{i,k}^n}(l_K) = \frac{1}{K}$, we get

$$l_K = \beta^2 r_0^{-2\alpha} \lambda^2 \log^{\frac{2}{t}} \frac{K r_0^2}{p^2}. \tag{II.29}$$

Now, we apply the following theorem by Uzgoren.

**Theorem II.1** (Uzgoren). *Let $x_1, \ldots, x_K$ be a sequence of i.i.d. positive random variables with continuous and strictly positive pdf $f_X(x)$ for $x > 0$ and cdf represented by $F_X(x)$. Let $h_X(x)$ be the growth function. Then, if $\lim_{x \to \infty} g'(x) = 0$, we have*

$$\log\left\{-\log F^K\left(l_K + h_X(l_K)u\right)\right\} = -u + \frac{u^2}{2!} h'_X(l_K) + \frac{u^3}{3!}\left(h_X(l_K) h''_X(l_K) - 2 h'^2_X(l_K)\right) + O\left(\frac{e^{-u + O(u^2 h'_X(l_K))}}{K}\right).$$

*Proof:* See [Uzg54, Equation 19] for proof. ∎

The above theorem gives taylor series expansion of the limiting distribution for Gumbel-type distributions. In particular, setting $l_K = \beta^2 r_0^{-2\alpha} \lambda^2 \log^{\frac{2}{t}} \frac{K r_0^2}{p^2}$ and $u = \log \log K$, we have $h(l_K) = O\left(\frac{1}{\log^{-\frac{2}{t}+1} K}\right)$, $h'(l_K) = O\left(\frac{1}{\log K}\right)$, $h''(l_K) = O\left(\frac{1}{\log^{\frac{2}{t}+1} K}\right)$, and so on. In particular, we have

$$Pr\left(\max_k \gamma_{i,k}^n \leq l_K + h(l_K) \log \log K\right) = e^{-e^{-\log \log K + O\left(\frac{\log^2 \log K}{\log K}\right)}} \tag{II.30}$$

$$= 1 - O\left(\frac{1}{\log K}\right), \tag{II.31}$$

where we have used the fact that $e^x = 1 + O(x)$ for small $x$. Similarly,

$$Pr\left(\max_k \gamma_{i,k}^n \leq l_K - h(l_K) \log \log K\right) = e^{-e^{\log \log K + O\left(\frac{\log^2 \log K}{\log K}\right)}} \tag{II.32}$$

$$= e^{-\left(1 + O\left(\frac{\log \log K}{\log K}\right)\right) \log K} \tag{II.33}$$

$$= O\left(\frac{1}{K}\right). \tag{II.34}$$

Subtracting (II.34) from (II.31), we get

$$Pr\left(l_K - \log \log K < \max_k \gamma_{i,k}^n \leq l_K + \log \log K\right) \geq 1 - O\left(\frac{1}{\log K}\right). \tag{II.35}$$

Note that the above equation is the same as (I.50). Therefore, following (I.51)-(I.61), we get

$$\mathcal{C}^* = BN\, O\left(\log\log^{2/t} \frac{Kr_0^2}{p^2}\right), \text{ and} \tag{II.36}$$

$$\mathcal{C}^* = BN f_{\text{lo}}^{\text{DN}}(r,B,N)\, \Omega\left(\log\log^{2/t} \frac{Kr_0^2}{p^2}\right). \tag{II.37}$$

Further, if $\frac{\log^{2/t} KN}{N} \gg 1$, then $\mathcal{C}^* = O\left(\log \frac{\log^{2/t} \frac{KN r_0^2}{p^2}}{N}\right)$.

## C. LogNormal

In this case, $|\nu_{i,k}^n|$ is distributed according to LogNormal-$(a,w)$ distribution. Hence, $|\nu_{i,k}^n|^2$ is distributed according to LogNormal-$(2a, 4w)$ distribution. The cumulative distribution function of $\gamma_{i,k}^n$, i.e., $F_{\gamma_{i,k}^n}(\gamma)$ (when $\gamma \geq 0$) is

$$F_{\gamma_{i,k}^n}(\gamma) = \int p\left(|\nu_{i,k}^n|^2 \leq \frac{\gamma}{g}\right) f_{G_{i,k}}(g) dg \tag{II.38}$$

$$= 1 - \frac{1}{2} \int_{\beta^2(p+d)^{-2\alpha}}^{\beta^2 r_0^{-2\alpha}} \text{erfc}\left[\frac{\log\frac{\gamma}{g} - 2a}{\sqrt{8w}}\right] f_{G_{i,k}}(g) dg, \tag{II.39}$$

where $\text{erfc}[\cdot]$ is the complementary error function. Using the asymptotic expansion of $\text{erfc}[\cdot]$, $F_{\gamma_{i,k}^n}(\gamma)$ can be approximated [AS70, Eq. 7.1.23] in the large $\gamma$-regime as:

$$F_{\gamma_{i,k}^n}(\gamma) \approx 1 - \frac{1}{2} \int_{\beta^2(p+d)^{-2\alpha}}^{\beta^2 r_0^{-2\alpha}} f_{G_{i,k}}(g) \frac{e^{-\left(\frac{\log\frac{\gamma}{g}-2a}{\sqrt{8w}}\right)^2}}{\left(\frac{\log\frac{\gamma}{g}-2a}{\sqrt{8w}}\right)\sqrt{\pi}} \sum_{m=0}^{\infty} (-1)^m \frac{(2m-1)!!}{2^m \left(\frac{\log\frac{\gamma}{g}-2a}{\sqrt{8w}}\right)^{2m}} dg, \tag{II.40}$$

where $(2m-1)!! = 1 \times 3 \times 5 \times \ldots \times (2m-1)$. We ignore the terms $m=1,2,\ldots$ as the dominant term for large $\gamma$ corresponds to $m=0$. Therefore, we have

$$F_{\gamma_{i,k}^n}(\gamma) = 1 - \sqrt{\frac{2w}{\pi}} \int_{\beta^2(p+d)^{-2\alpha}}^{\beta^2 r_0^{-2\alpha}} \frac{e^{-\left(\frac{\log\frac{\gamma}{g}-2a}{\sqrt{8w}}\right)^2}}{\log\frac{\gamma}{g}-2a} f_{G_{i,k}}(g) dg \tag{II.41}$$

$$= 1 - \sqrt{\frac{2w}{\pi}} \frac{r_0^2}{p^2} \frac{e^{-\left(\frac{\log\frac{\gamma}{\beta^2 r_0^{-2\alpha}}-2a}{\sqrt{8w}}\right)^2}}{\log\frac{\gamma}{\beta^2 r_0^{-2\alpha}}-2a} - \sqrt{\frac{2w}{\pi}} \int_{\frac{\beta^2}{(p-d)^{2\alpha}}}^{\frac{\beta^2}{r_0^{2\alpha}}} \frac{1}{\alpha\beta^2 p^2}\left(\frac{g}{\beta^2}\right)^{-1-\frac{1}{\alpha}} \frac{e^{-\left(\frac{\log\frac{\gamma}{g}-2a}{\sqrt{8w}}\right)^2}}{\log\frac{\gamma}{g}-2a} dg$$

$$+ \sqrt{\frac{2w}{\pi}} \int_{\frac{\beta^2}{(p+d)^{2\alpha}}}^{\frac{\beta^2}{(p-d)^{2\alpha}}} \frac{e^{-\left(\frac{\log\frac{\gamma}{g}-2a}{\sqrt{8w}}\right)^2}}{\log\frac{\gamma}{g}-2a} ds(g). \tag{II.42}$$

Now, we claim that

$$\lim_{\gamma \to \infty} \left(1 - F_{\gamma_{i,k}^n}(\gamma)\right)\left(\log\gamma - \log(\beta^2 r_0^{-2\alpha}) - 2a\right) e^{\left(\frac{\log\frac{\gamma}{\beta^2 r_0^{-2\alpha}}-2a}{\sqrt{8w}}\right)^2} = \frac{r_0^2}{p^2}\sqrt{\frac{2w}{\pi}}. \tag{II.43}$$

This is because the contribution of the two integrals in (II.42) towards the RHS of (II.43) is zero. The contribution of first integral, when $\gamma$ is large, is

$$\left| \left( \log \frac{\gamma}{\beta^2 r_0^{-2\alpha}} - 2a \right) e^{\left( \frac{\log \frac{\gamma}{\beta^2 r_0^{-2\alpha}} - 2a}{\sqrt{8w}} \right)^2} \int_{\frac{\beta^2}{(p-d)^{2\alpha}}}^{\frac{\beta^2}{r_0^{2\alpha}}} \frac{1}{\alpha \beta^2 p^2} \left( \frac{g}{\beta^2} \right)^{-1-\frac{1}{\alpha}} \frac{e^{-\left( \frac{\log \frac{\gamma}{g} - 2a}{\sqrt{8w}} \right)^2}}{\log \frac{\gamma}{g} - 2a} dg \right|$$

$$\leq \left( \log \frac{\gamma}{\beta^2 r_0^{-2\alpha}} - 2a \right) \frac{r_0^{-2\alpha-2}}{\alpha \beta^2 p^2} \int_{\frac{\beta^2}{(p-d)^{2\alpha}}}^{\frac{\beta^2}{r_0^{2\alpha}}} \frac{e^{\left( \frac{\log \frac{\gamma}{\beta^2 r_0^{-2\alpha}} - 2a}{\sqrt{8w}} \right)^2 - \left( \frac{\log \frac{\gamma}{g} - 2a}{\sqrt{8w}} \right)^2}}{\log \frac{\gamma}{g} - 2a} dg \quad \text{(II.44)}$$

$$\leq \frac{r_0^{-2\alpha-2}}{\alpha \beta^2 p^2} \int_{\frac{\beta^2}{(p-d)^{2\alpha}}}^{\frac{\beta^2}{r_0^{2\alpha}}} e^{\left( \frac{\log \frac{\gamma}{\beta^2 r_0^{-2\alpha}} - 2a}{\sqrt{8w}} \right)^2 - \left( \frac{\log \frac{\gamma}{g} - 2a}{\sqrt{8w}} \right)^2} dg \quad \text{(II.45)}$$

$$\leq \frac{r_0^{-2\alpha-2}}{\alpha \beta^2 p^2} \int_{\frac{\beta^2}{(p-d)^{2\alpha}}}^{\frac{\beta^2}{r_0^{2\alpha}}} e^{\frac{1}{8w} \left( \log \frac{\gamma^2}{g \beta^2 r_0^{-2\alpha}} - 4a \right) \log \frac{g}{\beta^2 r_0^{-2\alpha}}} dg \quad \text{(II.46)}$$

$$= \frac{r_0^{-2\alpha-2}}{\alpha \beta^2 p^2} \int_{\frac{\beta^2}{(p-d)^{2\alpha}}}^{\frac{\beta^2}{r_0^{2\alpha}}} \left( \frac{g}{\beta^2 r_0^{-2\alpha}} \right)^{\frac{1}{8w} \left( \log \frac{\gamma^2}{g \beta^2 r_0^{-2\alpha}} - 4a \right)} dg \quad \text{(II.47)}$$

$$\leq \frac{r_0^{-2\alpha-2}}{\alpha \beta^2 p^2} \int_{\frac{\beta^2}{(p-d)^{2\alpha}}}^{\frac{\beta^2}{r_0^{2\alpha}}} \left( \frac{g}{\beta^2 r_0^{-2\alpha}} \right)^{\frac{1}{8w} \left( \log \frac{\gamma^2}{\beta^4 r_0^{-4\alpha}} - 4a \right)} dg \quad \text{(II.48)}$$

$$= \frac{r_0^{-2\alpha-2}}{\alpha \beta^2 p^2} \frac{1}{\frac{1}{8w} \left( \log \frac{\gamma^2}{\beta^4 r_0^{-4\alpha}} - 4a \right)} \left( 1 - \left( \frac{r_0}{p-d} \right)^{\frac{2\alpha}{8w} \left( \log \frac{\gamma^2}{\beta^4 r_0^{-4\alpha}} - 4a \right) - 2\alpha} \right) \quad \text{(II.49)}$$

$$\to 0, \text{ as } \gamma \to \infty. \quad \text{(II.50)}$$

where in (II.44), we take an upper bound by taking the term $\left( \frac{g}{\beta^2} \right)^{-1-1/\alpha}$ out of the integral, and in (II.48), we put $g = \beta^2 r_0^{-2\alpha}$ in the exponent of $\left( \frac{g}{\beta^2 r_0^{-2\alpha}} \right)$ since $g \leq \beta^2 r_0^{-2\alpha}$. The second integral has an exponent term that goes to zero faster than $e^{-\left( \frac{\log \frac{\gamma}{\beta^2 r_0^{-2\alpha}} - 2a}{\sqrt{8w}} \right)^2} \to 0$, making its contribution zero. Note that only the first two term in (II.42) contribute to the RHS in (II.43). Similar to the above analysis, it is easy to show that

$$\lim_{\gamma \to \infty} f_{\gamma_{i,k}^n}(\gamma) \gamma e^{\left( \frac{\log \frac{\gamma}{\beta^2 r_0^{-2\alpha}} - 2a}{\sqrt{8w}} \right)^2} = \frac{r_0^2}{p^2 \sqrt{8w\pi}}. \quad \text{(II.51)}$$

Using the above equation and (II.43), we have

$$h(\gamma) = \frac{1 - F_{\gamma_{i,k}^n}(\gamma)}{f_{\gamma_{i,k}^n}(\gamma)} \approx \frac{4w\gamma}{\log \gamma} \text{ for large } \gamma, \text{ and} \quad \text{(II.52)}$$

$$\lim_{\gamma \to \infty} h'(\gamma) = 0. \quad \text{(II.53)}$$

Therefore, the limiting distribution of $\max_k \gamma_{i,k}^n$ belongs to the Gumbel-type. Solving for $l_K$, we have

$$l_K = \beta^2 r_0^{-2\alpha} e^{\sqrt{8w \log \frac{K r_0^2}{p^2} + \Theta(\log \log K)}}, \text{ and} \quad \text{(II.54)}$$

$h(l_K) = O\left(\frac{l_K}{\log l_K}\right)$, $h'(l_K) = O\left(\frac{1}{\log l_K}\right)$, $h''(l_K) = O\left(\frac{1}{l_K \log l_K}\right)$, and so on. Using Theorem II.1 for $u = \log \log K$, we have

$$Pr\left(\max_k \gamma_{i,k}^n \leq l_K + h(l_K) \log \log K\right) = e^{-e^{-\log \log K + O\left(\frac{\log^2 \log K}{\sqrt{\log K}}\right)}} \quad \text{(II.55)}$$

$$= 1 - O\left(\frac{1}{\log K}\right), \quad \text{(II.56)}$$

where we have used the fact that $e^x = 1 + O(x)$ for small $x$. Similarly,

$$Pr\left(\max_k \gamma_{i,k}^n \leq l_K - h(l_K) \log \log K\right) = e^{-e^{\log \log K + O\left(\frac{\log^2 \log K}{\sqrt{\log K}}\right)}} \quad \text{(II.57)}$$

$$= e^{-\left(1 + O\left(\frac{\log \log K}{\sqrt{\log K}}\right)\right) \log K} \quad \text{(II.58)}$$

$$= O\left(\frac{1}{K}\right). \quad \text{(II.59)}$$

Combining (II.56) and (II.59), we get

$$Pr\left(l_K - c\frac{e^{\sqrt{8w \log K}}}{\sqrt{\log K}} \log \log K < \max_k \gamma_{i,k}^n \leq l_K + c\frac{e^{\sqrt{8w \log K}}}{\sqrt{\log K}} \log \log K\right) \geq 1 - O\left(\frac{1}{\log K}\right), \quad \text{(II.60)}$$

where $c$ is a constant. Now, following a similar analysis as in (I.51)-(I.61), we get

$$\max_k \gamma_{i,k}^n = \Theta(l_K) \text{ w.h.p.}, \quad \text{(II.61)}$$

$$\mathcal{C}^* = BN\, O\left(\sqrt{\log \frac{Kr_0^2}{p^2}}\right), \text{ and} \quad \text{(II.62)}$$

$$\mathcal{C}^* = BN f_{\text{lo}}^{\text{DN}}(r, B, N)\, \Omega\left(\sqrt{\log \frac{Kr_0^2}{p^2}}\right). \quad \text{(II.63)}$$

Further, if $\frac{e^{\sqrt{\log KN}}}{N} \gg 1$, then $\mathcal{C}^* = O\left(\log \frac{e^{\sqrt{\log \frac{KNr_0^2}{p^2}}}}{N}\right)$.

## III. Proof of Theorem 4 and Corollary 3 for regular extended networks

The proof of Theorem 4 and Corollary 3 under the setup of regular extended networks is similar to the proof of Theorem 2 and Corollary 1 with the change of replacing $p^2$ by $BR^2$. Here, we use Lemma I.2, and Lemma I.3 with $p^2 = BR^2$. Lemma I.1, however, is replaced by Lemma III.1 (given below) wherein we show that the interference term in SNR expression can be bounded by a constant for extended networks.

We now give the proof details. The counterpart of Lemma I.1, i.e., Lemma III.1, for regular extended networks is as follows.

**Lemma III.1.** *In a regular extended-network and Rayleigh fading channel, i.e., $\nu_{i,k}^n \sim \mathcal{CN}(0,1)$, the lower bound on achievable sum-rate of the system is*

$$\mathcal{C}^* \geq f_{\text{lo}}^{\text{EN}}(r, N) \sum_{i,n} \mathrm{E}\left\{\log\left(1 + P_{\text{con}} \max_k \gamma_{i,k}^n\right)\right\}, \quad \text{(III.1)}$$

*where $r > 0$ is a fixed number, $f_{\text{lo}}^{\text{EN}}(r, N) = \frac{(1+r^2)^{-1}r^2}{N+(\mu+r\sigma)c_0}$, $\mu$ and $\sigma$ are the mean and standard-deviation of $|\nu_{i,k}^n|^2$, and $c_0 = \frac{P_{\text{con}}\beta^2 r_0^{2-2\alpha}}{R^2}\left(4 + \frac{\pi}{\sqrt{3}(2\alpha-2)}\right)$.*

*Proof:* We have, from Theorem 1,

$$\mathcal{C}^* \geq \sum_{i,n,k} \mathrm{E}\left\{\frac{\max_k \log\left(1 + P_{\mathsf{con}}\gamma_{i,k}^n\right)}{N + P_{\mathsf{con}}\beta^2 \sum_{j \neq i} R_{j,k}^{-2\alpha}|\nu_{j,k}^n|^2}\right\}, \tag{III.2}$$

where the expectation is over user-location set $(\boldsymbol{x}, \boldsymbol{y}) \triangleq \{(x_k, y_k) \,\forall\, k\}$, and the fading random-variable set $\boldsymbol{\nu} \triangleq \{\nu_{i,k}^n \,\forall\, i, k, n\}$. We will first consider the expectation w.r.t. $\boldsymbol{\nu}$ for a given $(\boldsymbol{x}, \boldsymbol{y})$. Then, $R_{j,k}$ is known for all $(j, k)$. We now apply one-sided variant of Chebyshev's inequality (also called Cantelli's inequality) to the term in the denominator, $\sum_{j \neq i} R_{j,k}^{-2\alpha}|\nu_{j,k}^n|^2$. Let the mean and standard deviation of $|\nu_{j,k}^n|^2$ be $\mu$ and $\sigma$. Note that $\boldsymbol{\nu}$ is independent of $(\boldsymbol{x}, \boldsymbol{y})$ by assumption. Therefore, the mean and standard deviation of $\sum_{j \neq i} R_{i,k}^{-2\alpha}|\nu_{j,k}^n|^2$ are $\mu \sum_{j \neq i} R_{j,k}^{-2\alpha}$ and $\sigma \sqrt{\sum_{j \neq i} R_{j,k}^{-4\alpha}}$, respectively. Hence, applying Cantelli's inequality, we have

$$Pr\Big(\sum_{j \neq i} |\nu_{j,k}^n|^2 > \mu \sum_{j \neq i} R_{j,k}^{-2\alpha} + r\sigma\sqrt{\sum_{j \neq i} R_{j,k}^{-4\alpha}}\Big) \leq \frac{1}{1+r^2}$$

$$\implies Pr\Big(\sum_{j \neq i} |\nu_{j,k}^n|^2 > (\mu + r\sigma) \sum_{j \neq i} R_{j,k}^{-2\alpha}\Big) \leq \frac{1}{1+r^2}$$

$$\implies Pr\Big(\sum_{j \neq i} |\nu_{j,k}^n|^2 \leq (\mu + r\sigma) \sum_{j \neq i} R_{j,k}^{-2\alpha}\Big) \geq \frac{r^2}{1+r^2}, \tag{III.3}$$

where $r > 0$ is a fixed number.

Now, we break the expectation in (III.2) into two parts — one with $\sum_{j \neq i} |\nu_{j,k}^n|^2 > (\mu + r\sigma) \sum_{j \neq i} R_{j,k}^{-2\alpha}$ and other with $\sum_{j \neq i} |\nu_{j,k}^n|^2 \leq (\mu + r\sigma) \sum_{j \neq i} R_{j,k}^{-2\alpha}$. We then ignore the first part to obtain another lower bound. Therefore, we now have

$$\mathcal{C}^* \geq \frac{r^2}{1+r^2} \sum_{i=1}^{B} \sum_{n=1}^{N} \frac{\mathrm{E}\left\{\max_k \log\left(1 + P_{\mathsf{con}}\gamma_{i,k}^n\right)\right\}}{N + (\mu + r\sigma)P_{\mathsf{con}}\beta^2 \sum_{j \neq i} R_{j,k}^{-2\alpha}}. \tag{III.4}$$

Applying coordinate geometry in order to upper bound $\sum_{j \neq i} R_{j,k}^{-2\alpha}$, the coordinate of any base-station can be written in the form of $m(2R, 0) + n(R, R\sqrt{3})$, where $m, n$ are integers, and $2R$ is the distance between neighbouring base-stations as shown in Fig. III.1.

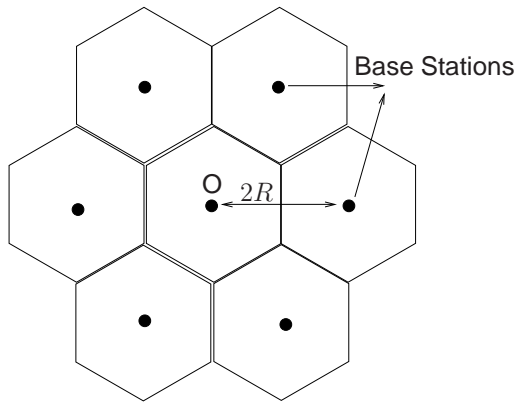

Fig. III.1: A regular extended network setup.

We can write

$$\sum_{j \neq i} R_{j,k}^{-2\alpha} = r_0^{-2\alpha} B_{\leq r_0} + \sum_{j \neq i, R_{j,k} > r_0} R_{j,k}^{-2\alpha}. \tag{III.5}$$

where $B_{\leq r_0}$ are the number of base-stations within $r_0$ distance of user $k$ with corrdinates $(x_k, y_k)$. Now,

$$\sum_{j \neq i, R_{j,k} > r_0} R_{j,k}^{-2\alpha} = \sum_{\substack{m,n \text{ s.t.} \\ R_{j,k} > r_0}} \left( \sqrt{(2mR + nR - x_0)^2 + (nR\sqrt{3} - y_0)^2} \right)^{-2\alpha}$$

$$\leq \iint_{\{d(x,y) > r_0\}} \frac{dy dx}{\underbrace{\left( \sqrt{(2xR + yR - x_0)^2 + (yR\sqrt{3} - y_0)^2} \right)^{2\alpha}}_{d(x,y)}}$$

$$= \iint_{\{x'^2 + y'^2 > r_0^2\}} \frac{1}{2\sqrt{3}R^2} \frac{dy' dx'}{(x'^2 + y'^2)^\alpha} \tag{III.6}$$

$$= \frac{1}{2\sqrt{3}R^2} \int_{r=r_0}^{\infty} \int_{\theta=0}^{2\pi} r \frac{1}{(r^2)^\alpha} dr d\theta \tag{III.7}$$

$$= \frac{\pi}{\sqrt{3}R^2} \frac{r_0^{2-2\alpha}}{2\alpha - 2}, \tag{III.8}$$

where in (III.6), we substituted $x' = 2xR + yR - x_0$ and $y' = yR\sqrt{3} - y_0$, and in (III.7), we substituted $x' = r\cos\theta$ and $y' = r\sin\theta$. Moreover, $B_{\leq r_0} \approx \frac{r_0^2}{R^2} < \frac{4r_0^2}{R^2}$. Therefore, from (III.5), we have

$$\sum_{j \neq i} R_{j,k}^{-2\alpha} < \frac{r_0^{2-2\alpha}}{R^2} \left( 4 + \frac{\pi}{\sqrt{3}(2\alpha - 2)} \right). \tag{III.9}$$

Combining the above equation with (III.4), we get

$$\mathcal{C}^* \geq \sum_{i,n} \mathrm{E} \left\{ \frac{\frac{r^2}{1+r^2} \max_k \log\left(1 + P_{\text{con}} \gamma_{i,k}^n\right)}{N + (\mu + r\sigma)c_0} \right\}, \tag{III.10}$$

where $c_0 = \frac{P_{\text{con}} \beta^2 r_0^{2-2\alpha}}{R^2} \left( 4 + \frac{\pi}{\sqrt{3}(2\alpha - 2)} \right)$ is a constant. From (III.10), we have

$$\mathcal{C}^* \geq f_{\text{lo}}^{\text{EN}}(r, N) \sum_{i,n} \mathrm{E} \left\{ \log\left(1 + P_{\text{con}} \max_k \gamma_{i,k}^n\right) \right\}, \tag{III.11}$$

where $f_{\text{lo}}^{\text{EN}}(r, N) = \frac{(1+r^2)^{-1} r^2}{N + (\mu + r\sigma)c_0}$. ∎

Note that, for Rayleigh fading channels,

$$|\nu_{i,k}^n|^2 \sim \text{exponential}(1) \ \forall \ i, k, n. \tag{III.12}$$

Therefore, $\mu = \sigma = 1$. The rest of the steps in the proof remain same as those in the proof of Theorem 2 and Corollary 1 in Section I with $p^2 = BR^2$. In particular, applying Lemma I.2, and Lemma I.3 for extended networks yield $l_K = \beta^2 r_0^{2\alpha} \log \frac{K r_0^2}{BR^2}$. Therefore, using [SH05, Theorem 1], we get

$$Pr\left\{ l_K - \log\log \frac{K}{B} \leq \max_k \gamma_{i,k}^n \leq l_K + \log\log \frac{K}{B} \right\} \geq 1 - O\left(\frac{1}{\log \frac{K}{B}}\right). \tag{III.13}$$

Note that the only change in the above equation from (I.50) is that $K$ is replaced by $K/B$. Following the analysis in (I.51)-(I.61) (after replacing $K$ by $K/B$), we get the final results in Theorem 4 and Corollary 3.

## IV. Proof of Theorem 5 and Corollary 4 for regular extended networks

For proof of this theorem and corollary, replace $p^2$ in the proof of Theorem 3 and Corollary 2 in Section II by $BR^2$ and use Lemma III.1 (instead of Lemma I.1) to obtain the lower and upper bounds for regular extended networks.